\documentclass{article}
\usepackage[margin=2cm]{geometry}
\usepackage[hyphens]{url}
\usepackage{graphicx}
\usepackage[numbers]{natbib}
\usepackage{caption}
\usepackage{comment}

\usepackage{booktabs}

\usepackage[linesnumbered,noend]{algorithm2e}
\usepackage{enumitem}
\usepackage{xcolor}
\usepackage{soul}
\usepackage{nicefrac}
\usepackage{amsmath,amssymb,amsfonts,amsthm}

\theoremstyle{definition}
\newtheorem{definition}{Definition}
\newtheorem{theorem}{Theorem}

\newtheorem{lemma}{Lemma}

\DeclareMathOperator{\periph}{periph}
\DeclareMathOperator{\rank}{rank}
\DeclareMathOperator{\Top}{top}

\definecolor{rankone}{HTML}{D9EAD3}
\definecolor{ranktwo}{HTML}{D9EAF7}
\definecolor{rankthree}{HTML}{FFF2CC}
\definecolor{rankfour}{HTML}{FCE5CD}
\definecolor{rankfive}{HTML}{EADCF8}

\newcommand{\prefclass}[2]{%
	\begingroup
	\setlength{\fboxsep}{2.5pt}%
	\colorbox{#1}{\ensuremath{\vphantom{b}#2}}%
	\endgroup
}

\newcommand{\rankone}[1]{\prefclass{rankone}{#1}}
\newcommand{\ranktwo}[1]{\prefclass{ranktwo}{#1}}
\newcommand{\rankthree}[1]{\prefclass{rankthree}{#1}}
\newcommand{\rankfour}[1]{\prefclass{rankfour}{#1}}
\newcommand{\rankfive}[1]{\prefclass{rankfive}{#1}}

		\title{
			\Large{Proportionality in Elections under Feasibility Constraints and Ranked Ballots:\\ The Solid Coalition Refinement Rule}
		}
		\usepackage{authblk}

		\author{
			Grzegorz Kwacz$^1$\quad
			Georgios Papasotiropoulos$^1$\quad
			Piotr Skowron$^{1}$ \\
			\vspace{0.1cm}
			{\small
				$^1$\textit{University of Warsaw}
			}
		}
		\date{}
		\allowdisplaybreaks
		\usepackage{cleveref}

		\Crefname{definition}{Definition}{Definitions}
		\Crefname{theorem}{Theorem}{Theorems}
		\Crefname{lemma}{Lemma}{Lemmas}
		\Crefname{corollary}{Corollary}{Corollaries}
		\usepackage[T1]{fontenc}

\begin{document}

			\maketitle

			\begin{abstract}
				We study elections with ranked ballots and general feasibility constraints, which indicate which sets of candidates are allowed as outcomes. We introduce the first proportionality axiom for this general setting and show that it extends the established notions from specific models that fit our general framework. We extend the Solid Coalition Refinement (SCR) rule to general constraints and prove that it satisfies our axiom exactly for matroid constraints. Although the naive implementation of SCR involves a step that is computationally intractable for weak orders, we show that in practical cases it can be implemented in polynomial time or efficiently in FPT time. We also illustrate limitations of this rule for approval ballots and provide its first empirical evaluation.
			\end{abstract}

			\section{Introduction}

			The formal model of an \emph{election} involves two largely independent choices. First, one specifies how voters express their preferences. This can be done through approval ballots, numerical scores, or rankings. Second, one specifies what qualifies as a valid outcome. In a committee election, this may mean selecting at most a fixed number of candidates. In participatory budgeting, the chosen projects must fit within the available budget. When several issues are decided simultaneously, the outcome may need to contain exactly one option for each issue. Some combinations of alternatives may also be ruled out because they are incompatible. We refer to all such requirements as \emph{feasibility constraints}. We study the setting in which \emph{voters submit ranked ballots and the outcome is subject to arbitrary such constraints}. By doing so, we take a step toward a general theory of fairness in voting.
			We formalize fairness through \emph{proportionality}: if a group comprising an $\alpha$-fraction of the electorate has sufficiently cohesive preferences, it should influence approximately an $\alpha$-fraction of the outcome.\looseness-1

			\medskip
			\noindent\textbf{Ranked ballots.}
			Rankings are a fundamental language for expressing preferences: each voter submits a strict or weak order over all or a subset of the candidates. They are used in elections for national legislatures and local governments \cite{rankedwiki}, as well as in participatory budgeting \cite{pbwiki}. Controlled experiments suggest that citizens prefer rankings to other ballot formats \cite{yang2024designing}. Beyond politics, ranked preferences underlie matching markets, including school and internship assignment, some Academy Award procedures \cite{oskars}, search and recommendation systems, AI alignment, and expert selection processes.\looseness-1

			\medskip
			\noindent\textbf{Feasibility constraints.}
			A feasibility system specifies which subsets of candidates may be selected, thereby abstracting away from the details of a particular application. It can represent selecting a committee of bounded size, choosing one alternative for each of several issues, or funding projects subject to a budget constraint. It can encode logical consistency requirements in judgment aggregation and restrictions such as demographic diversity bounds, regional quotas in participatory budgeting, or limits on the number of selections from each given category. It captures settings where some candidates are required to be selected. Such constraints arise in many real-world elections and are explicitly specified in their bylaws and regulations. For example, the AAAI Executive Council requires fewer than half of its members to have potential conflicts of interest; the Green parties in Germany and Poland impose gender quotas; PB processes restrict the joint selection of conflicting projects (e.g., in Katowice) or impose neighbourhood-specific budget limits (e.g., in Amsterdam); World Athletics Council elections impose gender and geographical constraints; and CoARA Steering Board elections additionally impose quotas on the representation of different types of organizations, such as universities, research centres, and funding organizations.

			\medskip
			Each of these two dimensions is already well understood when the other is restricted. In particular, there is a developed theory of fairness for arbitrary feasibility constraints combined with approval ballots, i.e., dichotomous rankings~\cite{masavrik2024generalised}. Similarly, there is already quite a good understanding of fairness for ranked ballots combined with a simple cardinality constraint, i.e., for committee elections.
			Combining ranked ballots with arbitrary feasibility constraints, however, poses a qualitatively different challenge.

			Methods developed for approval preferences under general constraints do not extend directly to rankings; nor do the more recent methods designed for numerical scores~\cite{skowron2026general}. Nevertheless, these methods typically follow a common template: they construct the outcome sequentially and, at each step, disregard candidates whose addition to the current partial solution would violate the feasibility constraints. This observation suggests a natural approach to our problem: start from a committee-monotonic rule for ranked ballots, i.e., a rule whose outcome for a committee size $x$ is a superset of its outcome for a committee size $y$ for any $x>y$, and adapt its selection process to accommodate general feasibility constraints.
			This approach, however, need not lead to a satisfactory outcome. In particular, feasibility cannot simply be imposed after first computing a proportional ranking. To see this, consider an election over the candidates $\{c_i,c_i' : i\in[m]\}$, subject to the constraint that exactly one candidate from each pair $\{c_i,c_i'\}$ must be selected. Suppose that $51\%$ of the voters rank
			\(
			c_1\succ c_2 \succ \dots \succ c_m \succ c_1' \succ c_2' \succ \dots \succ c_m',
			\)
			whereas the remaining $49\%$ rank
			\(
			c_1' \succ c_2' \succ \dots \succ c_m' \succ c_1 \succ c_2 \succ \dots \succ c_m.
			\)
			Reverse-sequential variants of Single Transferable Vote and the Expanding Approvals Rule~\cite{aziz2025committee}---committee-monotonic rules satisfying strong proportionality guarantees---would produce the alternating ranking
			\(
			c_1 \succ c_1' \succ c_2 \succ c_2' \succ \dots \succ c_m \succ c_m'.
			\)
			This ranking represents the two groups almost perfectly: candidates preferred by the two groups appear in alternating positions. However, scanning the ranking from top to bottom and selecting each candidate whenever feasibility permits yields the outcome $\{c_i\}_{i\in[m]}$, because, once $c_i$ is selected, its counterpart $c_i'$ becomes infeasible, and hence almost half of the electorate receives no representation at all.

			Thus, preferences and feasibility must be handled jointly, rather than in two separate stages. We show that another committee-monotonic rule for ranked preferences---the Solid Coalition Refinement (SCR) rule~\cite{aziz2025committee}---can be successfully extended to arbitrary feasibility constraints.

			\subsection{Contribution}
			We introduce the first proportionality axiom for ranked voting under general outcome constraints. Our axiom combines ideas from the proportionality axioms developed for ranked ballots in committee elections \cite{aziz2025committee} and approval ballots with feasibility constraints \cite{masavrik2024generalised}.
			We show that our axiom generalizes the established properties in these two restricted settings.

			We provide a voting rule with provable guarantees for our setting, which is obtained as a natural extension of the Solid Coalition Refinement rule \cite{aziz2025committee}.
			We show that it satisfies our proportionality notion whenever the outcome constraints form a matroid, a class that captures multiple natural and important relevant applications such as committee elections, public decision making over multiple issues and elections with disjoint candidate attributes. Complementing this result, we construct a family of instances showing that, for every non-matroid structure, there exists an election in which the rule fails to produce a proportional outcome.

			We also present a comprehensive set of findings on the computational complexity of the proposed rule. Although SCR is polynomial-time computable for strict rankings (and continues to be so under general outcome constraints), it becomes intractable for weak rankings unless $\mathrm{P}=\mathrm{coNP}$ \cite{aziz2025committee,brill2023robust}. We identify the first natural parameters for weak rankings where SCR admits fixed-parameter tractable algorithm, even in the presence of general constraints.
			The rule becomes tractable for two preference levels (approval ballots), whenever the equivalence classes in the weak rankings are bounded (a relaxation of strict rankings that is particularly practically relevant~\cite{mccune2024}), and for elections with a small number of voters.

			Finally, motivated by the tractability of SCR for approval ballots, we take a closer look at this special case and provide a more fine-grained analysis of its fairness guarantees in the committee elections setting.
			We study the proportionality degree~\cite{skowron2021proportionality} of SCR and show that it can be surprisingly low.
			We prove this result by first constructing a tailored hand-crafted instance and then using computer-aided instance generation to strengthen it---a technique that may be of independent interest.
			An additional limitation is that the rule fails to guarantee Pareto optimality.

			We complement our theoretical findings with the first empirical analysis of SCR. We use real-world datasets that cover the tractable cases we identified.

			\subsection{Related Work}
			The question of proportional representation under general feasibility constraints has only recently begun to receive attention. For the approval setting, \citet{masavrik2024generalised} showed how PAV and Phragmén's rule can be extended from committee elections to arbitrary constraints and proved that these rules satisfy a generalizations of the well-known EJR \cite{aziz2017justified} and PJR \cite{SANCHEZFERNANDEZ2026104503} axioms, respectively, if and only if the given constraints forms a matroid. PAV is also considered by \citep{mavrov2023fair}, who showed that yet another generalization of EJR (see \citep{masavrik2024generalised} for a comparison of the two notions), called restrained-EJR, is satisfied when the constraints are independent sets of a matroid. Beyond approval ballots, this work also considers cardinal ballots (i.e., where each voter assigns a numerical value to each candidate), aiming for approximate versions of the cardinal-utility-centric core stability property \citep{Droop1881,Fain2018,Lindahl1958}. Proportionality aspects for elections with general feasibility constraints under cardinal ballots have also been considered in a recent work by \cite{skowron2026general}, where adaptations of MES \cite{peters2020proportionality}, Phragmén's rule \cite{BrillFJL17phragmen}, and BOS \cite{papasotiropoulos2024method} to the general setting are introduced and proportionality guarantees of these rules are established.

			For ranked ballots, within the framework of committee elections, the classic proportionality notion is \emph{proportionality for solid coalitions} (PSC), introduced by \citet{Dummett1984} and extended to weak rankings (IPSC) by \citet{aziz2021proportionally}; quantifiable versions of PSC have been proposed by \citet{bardal2025proportional}. These properties guarantee representation proportional to the size of groups of voters that rank the same block of candidates above all others, irrespective of their internal ordering. PSC has been described as an absolutely essential condition for proportionality \citep{woodall1994properties} and is satisfied by rules including STV \cite{tideman1995single,woodall1994properties}, and the Expanding Approvals Rule \cite{aziz2020expanding}. The Solid Coalition Refinement rule was very recently proposed by \citet{aziz2025committee} as the first rule satisfying both PSC and committee monotonicity—the property that electing $k$ candidates should yield a superset of the winners for any smaller committee size. A candidate-monotone probabilistic voting rule based on Phragmén's Ordered Rule \cite{janson2016phragmen} that also satisfies PSC in expectation was proposed by \citet{peters2025candidate}. Strengthenings of IPSC have been proposed by \citet{brill2023robust}, such as rank-PJR+, which is satisfied by EAR but not by STV. Other stronger notions of proportionality include local stability~\cite{localStability} and the core~\cite{jiang2019approx,cheng2019group,pier-sko:core-stable}, yet these properties are not always satisfiable.
			Alternative notions that focus on pairwise comparisons between input and output rankings have been considered by \citet{lederer2025proportional} and~\citet{lederer2024squared}.\looseness-1

			Over the recent years proportionality has become a central theme of social choice, with applications extending well beyond political elections~\cite{halpern2026representation,kellerhals2024proportional,papasotiropoulos2026fairness}.

			\section{Election Setting}

			An election under \emph{ranked voting} is a quadruple $E = (C, N, \mathcal F, \mathcal R)$, where $C = \{c_1, c_2, \ldots, c_m\}$ is a set of $m$ \emph{candidates}, $N = \{1, 2, \ldots, n \}$ is a set of $n$ \emph{voters}, $\mathcal F \subseteq 2^C$ is a nonempty family of \emph{feasible outcomes}, and
			$\mathcal R = (\succeq_1, \succeq_2, \ldots, \succeq_n)$ is a profile of weak orders over $C$, representing voters' preferences.
			If the feasible outcomes consist of exactly sets of size at most $k$, for some integer $k$, we refer to this setting as \emph{committee elections}.
			If no voter is indifferent between any two candidates, we classify their ballots as \emph{strict rankings}.
			If each voter's ranking has at most two tiers, so called dichotomous rankings, we refer to the corresponding setting as \emph{approval voting}. In the latter case we denote by $A_i$ the upper tier of voter $i$.
			A \emph{selection rule}, or simply a \emph{rule} $\mathcal{S}$ is a function mapping an election  $E = (C, N, \mathcal F, \mathcal R)$ to an outcome $\mathcal{S}(E)=W\in \mathcal F.$

			\section{Proportionality Guarantees}
			\label{sec:prop}

			In this section, we introduce our main notion for group fairness for the considered general setting of elections. We build on the notion of a \emph{solid coalition for weak rankings}, or, in short \emph{solid coalition}, originally termed \emph{generalized solid coalition} by \citet{aziz2020expanding}.

			\begin{definition}
				Given an election $E=(C,N,\mathcal{F},\mathcal{R}),$
				a solid coalition of $E$ is a pair $(S, X)$, where $S \subseteq N$ and $X \subseteq C$, such that every voter in $S$ weakly prefers each candidate in $X$ to each candidate in $C \setminus X$.
			\end{definition}

			Towards a proportionality axiom, we require that any solid coalition $(S, X)$ is represented in the outcome by candidates that the voters in $S$ find acceptable as substitutes for those in $X$. Specifically, a candidate $c \notin X$ is an \emph{acceptable substitute} if at least one voter in $S$ ranks $c$ no worse than some candidate in $X$. Following the terminology of~\citet{aziz2020expanding}, we capture this via the \emph{periphery} of $X$ with respect to $S$: the set of all candidates that some voter in $S$ considers at least as good as some candidate in $X$. This relaxation for the satisfaction of voters in $S$ is necessary for ensuring satisfiability of the corresponding axiom~\cite{aziz2020expanding}.

			\begin{definition}
				Given an election $E=(C,N,\mathcal{F},\mathcal{R}),$ and a solid coalition $(S, X)$ we define
				$\periph(S, X) = \{ c \in C: \exists i \in S, c' \in X$ such that $c \succeq_i c' \}$.
			\end{definition}

			We now define the axiom of \emph{Proportionality for Solid Coalitions for Election under Feasibility Constraints}, or, in short, \emph{Proportionality}. The periphery naturally sets the stage: a coalition $(S, X)$ is \emph{$\ell$-represented} in the outcome if at least $\ell$ members of $\periph(S, X)$ appear in $W$, capturing the idea that acceptable substitutes count toward the coalition's share.
			The subtle challenge is deciding when $S$ has a legitimate claim to $\ell$ representatives.
			We proceed in the spirit of the axioms introduced by \citet{masavrik2024generalised} to capture proportionality under general feasibility constraints in the approval voting setting.
			Granting $\ell$ representatives to $S$ could block other voters from candidates they care about, due to feasibility constraints.
			Take any $T \subseteq W$. Although arbitrary, it intuitively represents the candidates over which the voters in $N\setminus S$ can exercise their influence.
			If $T$ and $\ell$ candidates from $X$ can coexist in a feasible outcome, there is no conflict and the claim of $S$ can be guaranteed. If they cannot coexist, the claim remains legitimate as long as $S$ is proportionally large enough to outweigh the rest---formally, that the claim of $S$ to $\ell$ candidates is stronger, relative to its size, than the remaining voters' claim to $T$. This is captured by $\frac{|S|}{n} > \frac{\ell}{|T| + \ell}$, which is equivalent to $\frac{|S|}{\ell} > \frac{n-|S|}{|T|},$ whenever $|S| < n$ and $T \neq \emptyset$.
			A solid coalition \emph{deserves $\ell$ candidates} when no set $T$ forms valid evidence against its claim, and the axiom that we propose demands that every deserving coalition is indeed represented.

			\begin{definition}
				\label{def:prop}
				Given an election $E=(C,N,\mathcal{F},\mathcal{R}),$ we say that a solid coalition $(S, X)$ \emph{deserves $\ell$ candidates in $W$} if for every $T \subseteq W$ either there exists a set $X' \subseteq X$ with $|X'| \geq \ell$ such that $X' \cup T \in \mathcal F$ or
				\( \frac{|S|}{n} > \frac{\ell}{|T| + \ell}.\)
				We say that an outcome $W \in \mathcal F$ satisfies \emph{Proportionality} if for every positive integer $\ell$ and every solid coalition $(S, X)$ that deserves $\ell$ candidates in $W$ we have $|\periph(S, X) \cap W| \geq \ell$.
			\end{definition}

			To position our definition of proportionality within the literature, we provide the following findings. First, it extends \emph{generalized PSC}~\cite{aziz2020expanding} from committee elections to arbitrary feasibility constraints. The axiom called \emph{generalized Droop-PSC} is satisfied by a committee $W$ as long as for every natural number $\ell$ and for every solid coalition $(S,X)$ with $|S| > \ell \frac{n}{k+1}$, it holds that $W$ contains at least  $min(\ell, |X|)$ candidates from $\periph(S, X)$.

			\begin{theorem}
				\label{thm:drooppsc}
				Given a committee election under ranked voting $E=(C, N, \{ W' \subseteq C : |W'| \leq k \}, \mathcal R)$, an outcome $W \subseteq C$ with $|W| = k$ satisfies generalized Droop-PSC if and only if $W$ satisfies proportionality in $E$.
			\end{theorem}

			Second, the known equivalence~\cite{aziz2020expanding} from the committee election setting between generalized PSC and \emph{PJR} does not immediately carry over to the setting with feasibility constraints. PJR under feasibility constraints is defined by \citet{masavrik2024generalised} as follows.\looseness-1

			\begin{definition}
				Given an approval voting election $E = (C, N, \mathcal F, \mathcal A)$ and an outcome $W \in \mathcal F$, we say that a group of voters $S \subseteq N$ \emph{deserves $\ell$ candidates in $W$} if for each set $T \subseteq W$ either there exists a set $X' \subseteq \bigcap_{i \in S} A_i$ with $|X'| \geq \ell$ such that $T \cup X' \in \mathcal F$ or the following inequality holds:
				$\frac{|S|}{n} > \frac{\ell}{|T| + \ell}$.
				We say that an outcome $W \in \mathcal F$ of an election $E = (C, N, \mathcal F, \mathcal A)$ satisfies \emph{Proportional Justified Representation} if for every group of voters $S \subseteq N$ that deserves $\ell$ candidates we have $\left|\left(\bigcup_{i \in S}A_i\right) \cap W\right| \geq \ell$.
			\end{definition}

			We show that Proportionality always implies PJR, while the reverse is not always true. In particular, we show that they are equivalent whenever all inclusion-wise maximal feasible outcomes have the same size (e.g., under matroid constraints; see \Cref{def:matroids}), and that this condition is also necessary in the sense that whenever it is violated, there exists an election in which PJR is strictly weaker than Proportionality.

			\begin{theorem}
				Given an approval voting election under feasibility constraints $E=(C, N, \mathcal F, \mathcal A)$, if an outcome $W \in \mathcal F$ satisfies Proportionality then $W$ satisfies PJR in $E$.
				\label{thm:pjr1}
			\end{theorem}

			\begin{theorem}
				Given a family of feasible outcomes $\mathcal F$, if each inclusion-wise maximal set $W \in \mathcal F$ is the same size, then for each election $E=(C,N,\mathcal F,\mathcal A)$ and an inclusion-wise maximal outcome $W$ satisfying PJR, $W$ also satisfies Proportionality. If there exist two inclusion-wise maximal outcomes with different sizes, then there exists an election $E=(C,N,\mathcal F, \mathcal A)$ and an inclusion-wise maximal outcome $W$ satisfying PJR but not Proportionality.
				\label{thm:pjr2}
			\end{theorem}

			\subsection{The SCR Rule under General Constraints}

			The \emph{Solid Coalition Refinement} (SCR) rule, introduced by~\citet{aziz2025committee} for committee elections, starts from the empty outcome and repeatedly selects the candidate that best addresses the strongest underrepresented coalition. The underrepresentation value of a solid coalition $(S, X)$ under the current committee $W$ quantifies how far $S$ is from being proportionally represented: the higher the value, the stronger the claim of $S$ to the next seat. This value is defined as $\rho(W, S, X):=\frac{|S|}{|W \cap \periph(S, X)| + 1}$.
			When the most underrepresented coalition is $(S, X)$, the rule adds a candidate from $X \setminus W$.
			If several such candidates exist, it must decide which one in particular to select. To do so, it applies a process of \emph{refinement}: the selection procedure is applied iteratively to the candidates in $X$, narrowing the set down until it contains only one candidate not in $W$, which is then added to the committee.
			We denote the committee formed after a number $k$ of selections of candidates by $W_k$.

			The only modification needed to extend the rule to (downward-closed) feasibility constraints is to ensure feasibility is maintained at every step.
			Moreover, each refinement step should ensure that at least one electable candidate survives.
			We call the resulting rule the \emph{Solid Coalition Refinement for Feasibility Constraints,} or, in short \emph{Solid Coalition Refinement} (SCR) rule.
			The pseudocode is given in~\Cref{alg:scr}, where we assume a deterministic tie-breaker ensuring that the $\arg\max$ function returns a single output.

						\Cref{thm:drooppsc,thm:pjr1,thm:pjr2} hold for arbitrary feasibility constraints. Henceforth we restrict attention to downward-closed constraints. However, this assumption is not restrictive as any family of feasible sets can be transformed into a downward-closed one without affecting our theoretical results: given a family $\mathcal{F}$, we define $\mathcal{F}'$ to consist of all sets that can be extended to some set in $\mathcal{F}$. By construction, $\mathcal{F}'$ is downward closed and can then be used within the SCR rule. Then, any outcome returned by SCR is already feasible under $\mathcal{F}$ since SCR returns an inclusion-wise maximal set.

			\begin{algorithm}
				\caption{Solid Coalition Refinement rule for elections under feasibility constraints.\vspace{-1cm}}
				\label{alg:scr}
				$W \gets \emptyset, C' \gets \{c\in C : \{c\}\in \mathcal F\}$

				\While{$C'\neq \emptyset$}{

					$D \gets C$

					\While{$|D \cap C'| > 1$}{
						$\Phi(\mathcal{R}, W, D) \gets \{(S, X) : (S, X) \text{ is a }$
						$\text{solid coalition, } X \subsetneq D \text{, } |X \cap C'| \geq 1 \}$

						$(S^*, X^*) \gets \arg\max_{(S, X) \in \Phi(\mathcal{R}, W, D)} \frac{|S|}{|W \cap \periph(S, X)| + 1}$

						$D \gets X^*$
					}

					$W \gets W \cup (D\cap C')$

					$C' \gets \{ c \in C \setminus W : W \cup \{c\} \in \mathcal F\}$
				}

				\Return $W$
			\end{algorithm}

			Our main axiomatic result (\Cref{scr-prop}) shows that its outcome is proportional, in the sense of \Cref{def:prop}, provided that the input feasibility constraints form a matroid.
			Note that the proportionality guarantees provided by
			\citet{masavrik2024generalised} are also contingent on the feasibility constraints forming a \emph{matroid}.

			\begin{definition}
				\label{def:matroids}
				A family of feasibility constraints $\mathcal{F}$ is a matroid if for each $X, Y \in \mathcal F$ such that $|X| < |Y|$, there exists $c \in Y \setminus X$ such that $X \cup \{c\} \in \mathcal F$.
			\end{definition}

			The intuition is that candidates carry similar weight in a matroid and if removing two or more candidates from an outcome can make room for some new candidate, then in fact removing just one of them suffices.
			As noted by \citet{masavrik2024generalised}, matroid constraints already capture several natural settings, including committee elections, public decisions, and elections with disjoint candidate attributes. That said, important frameworks fall outside this class and would require separate treatment, such as Participatory Budgeting~\cite{PBsurvey} or elections with both positive and negative ballots~\cite{kraiczy2026proportional}. We view these as particularly important future directions.\looseness-1

			\begin{theorem}
				\label{scr-prop}
				Given an election $E=(C,N,\mathcal{F},\mathcal{R})$ where $\mathcal{F}$ is a matroid, $\textit{SCR}(E)$ satisfies Proportionality.
			\end{theorem}

			\begin{proof}[Proof Sketch]
				The proof of the result is based on two technical lemmas.
				The first guarantees that SCR continuously protects coalitions admitting feasible representation: if, at the beginning of a step, they support a candidate that can feasibly be added to the committee, then, after the step, they cannot remain excessively underrepresented.

				\begin{lemma}\label{thm:scr-electable-prop}
					Given an election $E=(C,N,\mathcal{F},\mathcal{R}),$ where $\mathcal{F}$ is a matroid, after each step $k$ of the SCR rule, for every solid coalition $(S,X)$, if there exists $c \in X \setminus W_{k-1}$ such that $\{c\} \cup W_{k-1} \in \mathcal F$ then $\rho(W_k, S, X) \leq \frac{n}{k + 1}$.
				\end{lemma}

				To show \Cref{thm:scr-electable-prop} we build on the payment-scheme approach that was used for proving that SCR satisfies PSC for committee elections \cite{aziz2025committee}.
				While this lemma’s underrepresentation value suffices for committee elections, going beyond it requires an additional step.
				In particular, while the proof of \Cref{thm:scr-electable-prop} indeed largely follows the arguments of \citet{aziz2025committee}, the proof of \Cref{thm:lem2prop} is entirely novel---it is independent of any specific rule and establishes a general invariant whose preservation is sufficient for satisfying the axiom.
				According to \Cref{thm:scr-electable-prop}, SCR maintains this invariant ensuring that, after each selection, every deserving coalition can still achieve proportional representation.

				\begin{lemma}\label{thm:lem2prop}
					Given an election $E=(C,N,\mathcal{F},\mathcal{R}),$ where $\mathcal{F}$ is a matroid, let $W$ be the outcome of SCR in $E$. After each step $k$ of the SCR rule, for every solid coalition $(S,X)$ that deserves $\ell$ candidates in $W$, there exists a set $X_k \subseteq X \setminus W_k$ such that $|X_k| + |W_k \cap \periph(S, X)| \geq \ell$ and $X_k \cup W_k \in \mathcal F$.
				\end{lemma}

				\Cref{thm:lem2prop} guarantees that at every step $k$, for each solid coalition $(S,X)$ deserving $\ell$ candidates in the outcome of the rule (in the sense of \Cref{def:prop}), there exists a set $X_k$ that can still be elected and electing it will result in $(S,X)$ being \emph{$\ell$-represented}.
				From this, \Cref{scr-prop} follows directly, because, when the rule terminates, no further feasible candidate can be added, meaning that $(S, X)$ is already $\ell$-represented.
			\end{proof}

			We conclude the section by establishing tightness for \Cref{scr-prop} with respect to the considered families of constraints: moving beyond matroids leads to a negative result.

			\begin{theorem}
				For each non-matroid family of feasibility constrains $\mathcal{F}$, there is an election $E=(C,N,\mathcal{F},\mathcal{R}),$ where $SCR(E)$ fails Proportionality.
				\label{thm:non-matroid}
			\end{theorem}

			\section{Algorithmic Aspects of the SCR Rule}
			\label{sec:algs}
			SCR performs particularly well under strict preferences. Not only it achieves strong proportionality guarantees even in the presence of general feasibility constraints making it the first rule known to produce proportional outcomes in such a broad framework under ranked ballots, but it is also polynomial-time computable \cite{aziz2025committee}.
			However, once we move to weak rankings,
			the naive implementation of the rule involves a step that is computationally hard, already in the setting of committee elections \cite{aziz2025committee}.
			We address this issue by establishing parameters of the input ballots that admit a fixed-parameter tractable algorithms, even under general feasibility constraints.

			We note that the transition from committee elections to general feasibility constraints introduces no additional computational burden. In particular, this holds under the assumption that we have \emph{access to an oracle} that, given a subset of the candidates, determines whether it is feasible. Such an oracle can be implemented in polynomial time for most of the natural feasibility constraints, and we assume its existence for the results of this section. The only component whose polynomial-time implementability is not immediate is the maximization procedure in Steps 5--6 of \Cref{alg:scr}.
			It is this maximization problem that is arising in the refinement loop of SCR that is coNP-complete when weak rankings are allowed in the input ballots \cite{aziz2025committee,brill2023robust}.
			The intractability result relies on weak rankings with three preference levels with a number of candidates per level that is a function of $n$. This leaves open two regimes in which tractability may still be possible:
			\begin{itemize}[noitemsep]
				\item preference structures of two levels---approval ballots,
				\item indifference classes in voters' ballots of bounded size---the first step beyond strict rankings.
			\end{itemize}
			We show that both admit efficient algorithms. Then, we complement these positive results by studying the parameterized complexity of the rule with respect to the number of voters and candidates, obtaining additional tractability results.\looseness-1

			\medskip

			We begin with the setting of approval voting. For a solid coalition $(S,X)$, we distinguish two cases. In the first, $X \subseteq \bigcap_{i \in S} A_i$. For this, we show that the periphery of $X$ with respect to $S$ is not affected by the size or the members of $X$, so the rule only needs to consider singleton sets $X$, since their supersets would yield the same underrepresentation value. This makes the refinement step unnecessary, as any refinement sequence resulting in electing a candidate $c$ can skip the intermediate refinements.
			The second case is when there exists some voter $i \in S$ such that $X \not\subseteq A_i$. Then we have $\periph(S, X) = C$, so again this does not depend on the set $X$. Therefore, we can restrict coalitions of this kind to those that refine the current set by removing only a single candidate from it.
			Taken together, these two cases imply that the refinement simplifies to either immediately electing or removing a candidate. This decision is based on a maximization that we show that it can be performed in polynomial time via a reduction to a submodular optimization problem---an approach that resembles a previously used one by \citet{brill2023robust} for a conceptually different problem.\looseness-1

			\begin{theorem}
				\label{thm:app}
				In the approval voting setting, the outcome of the SCR rule can be computed in polynomial time.
			\end{theorem}

			The second type of coalitions, discussed above, and considered at every step of SCR in the approval setting (i.e., coalitions that refine the set by removing a candidate), arise when closely following the original definition of the rule by \citet{aziz2025committee} for weak orders over all candidates and applying it to the approval setting. However, such coalitions are not a natural object of interest in the standard approval-based framework, as voters are typically assumed to derive no satisfaction from candidates outside their approval ballots. Given that these coalitions are technical artifacts, and that they are naturally proportionally represented in any outcome when the feasibility constraints satisfy the prerequisites of \Cref{thm:pjr2}, it is reasonable to avoid considering this step in the algorithm. For approval ballots, we call the resulting rule \emph{Greedy-PJR+}, as it greedily optimizes an axiom called \emph{PJR+}~\cite{brill2023robust}. We refer to \Cref{sec:experiments} for a comparison of this rule with SCR.

			\medskip

			As mentioned, SCR is polynomial-time computable for strict rankings, i.e., when each preference level contains a single candidate. We now show that SCR remains efficiently computable whenever the size of each preference level is bounded.
			We also extend this algorithm to handle truncated ballots, i.e., ballots where there are candidates that have not been explicitly ranked, and which are assumed to belong to the last preference level, with indifference classes of bounded size.
			This is, for example, the case in the large recently compiled dataset of Scottish local elections \cite{mccune2024}; we use this in the empirical evaluation (\Cref{sec:experiments}).

			\begin{theorem}
				\label{thm:boundedties}
				The outcome of the SCR rule can be computed in FPT time parameterized by the maximum number of tied candidates in a non-last level of a voter’s ranking.
			\end{theorem}

			Following the modification of the rule for approval ballots that led to the introduction of Greedy-PJR+, the same insight motivates a re-consideration of the implementation of SCR for truncated ballots. In this variant, we restrict attention to coalitions $(S,X)$ such that every member of $S$ includes all members of $X$ in a preference level above their last one. We evaluate this variant as well in \Cref{sec:experiments}.

			Once in the realm of parameterized algorithms, it is natural to ask what happens when we take the basic instance parameters as our focus. The size of indeference classes is bounded by the number of candidates, so the algorithm from \Cref{thm:boundedties} is also FPT with respect to $m$. Parameterizing with respect to $n$ needs another approach, but again leads to FPT.\looseness-1

			\begin{theorem}
				The outcome of the SCR rule can be computed in FPT time parameterized by the number of voters.
				\label{thm:FPTsmallelections}
			\end{theorem}

			We conclude by highlighting that the parameterized algorithms from \Cref{thm:boundedties,thm:FPTsmallelections} are highly relevant in practice, as their parameters are typically small. In particular, to establish this claim, we examined all 63 \textsc{Preflib} datasets with ranked ballots, covering a wide range of applications. In over 50\%, at least one of the parameters for which we establish FPT was at most 10, and in over 70\%, at most 15.

			\section{Limitations of SCR under Approval Voting}
			\label{sec:app}
			For ranked ballots with general feasibility constraints, our adaptation of SCR is the first rule known to provide fairness guarantees. Focusing on committee elections, under ranked ballots, SCR achieves a prominent guarantee that is among the strongest in the literature. Restricting the focus further, in this section we aim to understand how it compares with established proportional methods in the special case of approval ballots, which can serve as a natural benchmark. The result of \Cref{thm:app} makes this setting a suitable domain for applying the rule, provided that its theoretical properties remain satisfactory. Unfortunately, in this section we show that this is not the case, as SCR may fall short both in terms of fairness and efficiency.
			This leads to the conclusion that when the flexibility and expressiveness gained from ranked ballots and, even more so, from general feasibility constraints is not necessary for the application at hand, existing approaches such as MES, PAV, and Phragmén appear to be better suited.

			The simplicity of the approval setting allows us to perform a fine-grained analysis of fairness.
			By studying its proportionality degree \cite{skowron2021proportionality}, we show that the outcome of SCR can be particularly weak in that regard, even for committee elections.
			For comparison, note that the proportionality degree of PAV is $\ell-1$, of Phragmen is $\nicefrac{\ell-1}{2}$ and of MES is between $\nicefrac{\ell-1}{2}$ and $\nicefrac{\ell+1}{2}$ \cite{lackner2023multi}.\looseness-1

			\begin{definition}
				Let $E$ be a committee election under approval ballots. A group of voters $V$ is $\ell$-cohesive in $E$ if $|V|\geq \ell\,\nicefrac{n}{k}$ and all voters in $V$ approve at least $\ell$ common candidates.
				Fix a function $f:\mathbb{N}\rightarrow\mathbb{R}$. A rule $\mathcal{S}$ for committee elections under approval ballots has proportionality degree $f$ if, for every election $E$ and every $\ell$-cohesive group of voters $V$, the outcome $W=\mathcal{S}(E)$ satisfies
				$\frac{1}{|V|}\sum_{i\in V}|A_i\cap W|\geq f(\ell).$
			\end{definition}

			We prove the negative result regarding the proportionality degree of SCR in two steps.
			The key idea is to start from an analytically tractable instance (\Cref{thm:propdegree}) and then refine it via a linear program (LP) to obtain a strictly stronger bound (\Cref{thm:propdegreecomp}).
			This technique constitutes a methodology of independent interest for the creation of negative instances and may be useful beyond the specific rule and the considered elections domain.
			We first construct an intuitively clean instance that suffices to establish the weak performance of SCR with respect to the considered metric, for arbitrary $k$.

			\begin{theorem}
				\label{thm:propdegree}
				For each $k \in \mathbb{N}_{\geq 4}$, there exists an $\ell \in \mathbb{N}_+$ and an approval committee election $E=(C, N, \{ W' \subseteq C : |W'| \leq k \}, \mathcal A)$ with an $\ell$-cohesive group of voters $S \subseteq N$ such that $|A_i \cap SCR(E)| \leq \ell \cdot
				\frac{1}{\lfloor\sqrt{k}\rfloor - 1},\forall i\in S
				$.
			\end{theorem}

			Now we set the question of whether the instance used in the proof of \Cref{thm:propdegree} to upper bound the proportionality degree of SCR can be further improved. We answer this in the affirmative using a computer-aided approach. We formulate an LP that takes a committee size $k$ as input and outputs an instance inspired by the construction in \Cref{thm:propdegree}, with the goal of tightening the bound. This method allows us to derive improved bounds for small values of $k$.
			Since in many natural applications of committee elections the intended committee size is reasonably small, this result is informative in practice, even though it does not provide a closed-form expression valid for arbitrary $k$ as \Cref{thm:propdegree} does.

			\begin{theorem}
				\label{thm:propdegreecomp}
				For $k\leq 400$, there exists an $\ell \in \mathbb{N}_+$ and an approval committee election $E=(C, N, \{ W' \subseteq C : |W'| \leq k \}, \mathcal A)$ with an $\ell$-cohesive group of voters $S \subseteq N$ such that $|A_i \cap SCR(E)| \leq \ell \cdot \frac{3}{k^{0.72}}, \forall i\in S$.
			\end{theorem}

			\section{Empirical Analysis of the SCR Rule}\label{sec:experiments}

			In this section, we present the first experimental evaluation of SCR. We focus on two highly important scenarios where its outcome can be computed efficiently as shown in \Cref{sec:algs}:

			(a) truncated ballots with strict rankings,

			(b) approval ballots.

\begin{figure*}
				\centering
				\includegraphics[width=\textwidth]{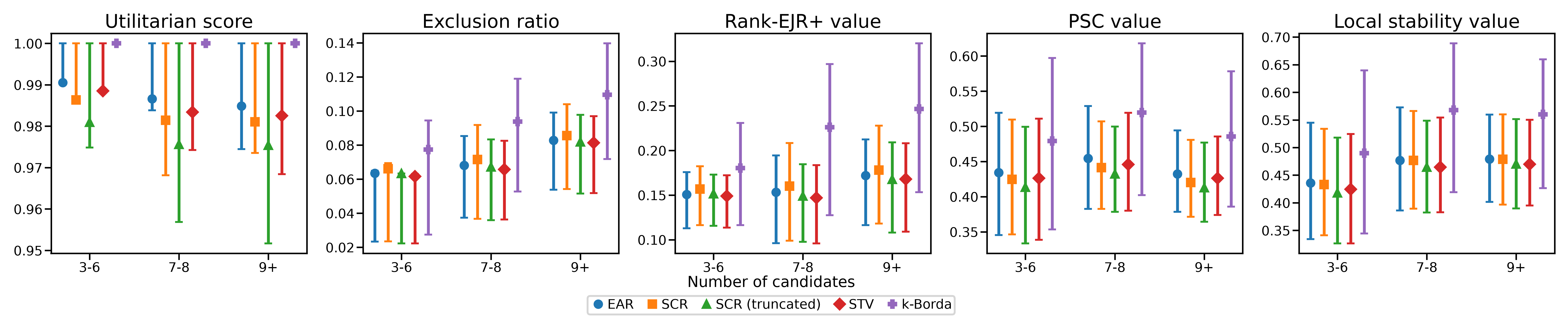}
				\includegraphics[width=\textwidth]{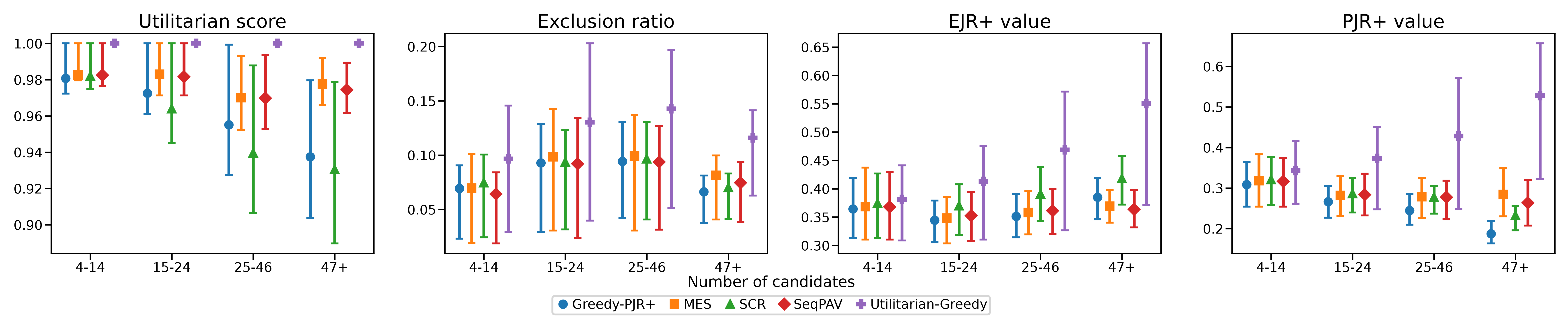}
				\caption{Average values of the evaluated metrics for the outcomes produced by the considered rules on strictly ordered truncated (\Cref{sec:exp_rank}) and approval ballots (\Cref{sec:exp_app}). Bars indicate the 25--75th percentile range.}
				\label{fig:all}
			\end{figure*}

			\noindent For each setting, we compare SCR against popular rules using suitable datasets. We conduct our comparison in the framework of committee elections: the most basic and well-studied feasibility constraint. This provides a clean experimental setting that allows for clear insights into the behavior of the rule, enabling direct comparison with the standard benchmarks studied in the relevant literature.
			Our main findings are presented in \Cref{fig:all}; additional results are provided in \Cref{sec:app_exp}.

			\subsection{Strictly Ordered Truncated Ballots}
			\label{sec:exp_rank}
			We use the dataset of Scottish local elections~\cite{mccune2024}, which consists of 1070 instances from local government elections held between 2007 and 2022. In these, voters were submitting strictly ordered truncated ballots.

			We compare SCR against the following rules: $k$-Borda, capturing efficiency-oriented approaches; and STV, one of the most established methods for ranked voting in real-life elections, including Scottish local elections; and EAR, a relatively recently introduced method with strong theoretical proportionality guarantees.
			We additionally consider a variant of SCR, denoted as \emph{truncated}, where a voter can belong to a coalition $(S,X)$ only if all candidates in $X$ are ranked on their ballot, i.e. they are not placed in the last preference level. We use an analogous interpretation for the implementation of the other rules.
			We refer the reader to \Cref{sec:app_exp} for the definitions of the considered rules.
			For the non-exhaustive rules (EAR and STV), we use the Add1 completion method~\citep{faliszewski2023pabulib}. Namely, when a rule returns an outcome of fewer than $k$ candidates, we repeat  after having slightly increased the voting power of each voter. We also tested other completion methods and
			Add1 consistently achieved better results, providing a stronger benchmark against which to evaluate SCR.
			The rules are evaluated based on the following metrics:

			\begin{description}[topsep=-1pt,itemsep=-1pt,leftmargin=*]
				\item[Utilitarian score:] We first assign Borda scores to ranked candidates of each voter, where the top-ranked candidate receives $m-1$ points, the second-ranked candidate receives $m-2$ points, and so on. Unranked candidates receive $0$ points. The total score of the committee from all voters is then normalized by the value achieved by the $k$-Borda rule.\looseness-1
				\item[Exclusion ratio:] This is a standard measure in the proportionality literature that captures the fraction of voters whose ballots do not contain any elected candidate. The dataset includes highly truncated ballots (specifically, only 13\% of ballots across all elections contain rankings over all candidates). Therefore, this metric captures the voters who did not rank any of the elected candidates.
				\item[PSC, Rank-EJR+ and Local stability Values:] These metrics are defined by~\citet{bardal2025proportional}, who introduce quantitative measures for several axioms in the ranked voting setting. They extend well-established binary metrics by asking for the smallest approximation factor of a group's size for which the axiom remains satisfied. Lower values indicate that the axiom is satisfied for a larger fraction of voter groups, and therefore correspond to stronger fairness guarantees.\looseness-1
			\end{description}

			With respect to utilitarian performance, SCR performs slightly worse than its competitors but still achieves a remarkable 97\% of the optimum on average. This modest utilitarian loss is traded for slightly stronger fairness guarantees. Across all fairness metrics and classes of instances considered, the truncated variant of SCR achieves comparable, sometimes even slightly better, performance to the state-of-the-art rules. It also offers marginal improvements against standard SCR. Although SCR bases its decisions on solid coalitions, which impose a rather strong requirement on groups of voters, the experiments demonstrate that this does not prevent the algorithm from making sound overall decisions.

			\subsection{Approval Ballots}
			\label{sec:exp_app}
			Given the particularly large number of instances, the diversity of voter preferences and election sizes, and the widespread use of \textsc{Pabulib}~\cite{faliszewski2023pabulib} in the relevant literature, we use its datasets for the approval voting setting. However, as provided, they do not directly fit the setting for which our axiomatic guarantees for SCR hold, since Participatory Budgeting does not correspond to matroid constraints. We therefore transform each instance into a committee election. We do so by setting $k$ equal to the budget divided by the average project cost (rounded to the nearest integer). We then discard the project costs and treat each project as a candidate competing for one of the $k$ committee seats.

			We compare SCR against Utilitarian Greedy Rule, MES, Greedy-PJR+, and Sequential PAV. We refer the reader to \Cref{sec:app_exp} for the definitions of these rules.

			Compared to the ranked setting (\Cref{sec:exp_rank}), the utilitarian performance of SCR deteriorates in the approval voting context. This is in line with the findings of \Cref{sec:app}. The gap is particularly pronounced in larger instances, where both MES and seqPAV outperform SCR to a notable extent. On Exclusion Ratio and PJR+, SCR loses its dominant position but remains competitive with the other state-of-the-art proportional rules, maintaining consistent performance across all instance sizes. For EJR+, SCR underperforms relative to its fairness-oriented competitors, although it remains substantially closer to them than to the utilitarian method.

			\vspace{-0.3cm}
			\section{Conclusion}
			Our results show that SCR provides strong fairness guarantees with general feasibility constraints. It is polynomial-time computable in several practically relevant cases.
			Nevertheless the rule performs worse than other well-established rules for approval ballots. This formulates an interesting open question, which we elaborate on in \Cref{sec:modified_solid_coalition_rule}.

			\vspace{0.3cm}
			{\small{\paragraph{Acknowledgements}
					The authors were supported by the European Union (ERC, PRO-DEMOCRATIC, 101076570). Views and opinions expressed are however those of the authors only and do not necessarily reflect those of the European Union or the European Research Council. Neither the European Union nor the granting authority can be held responsible for them.}}

			\vspace{-0.3cm}
			\begin{figure}[h]
				\includegraphics[width=0.2\linewidth]{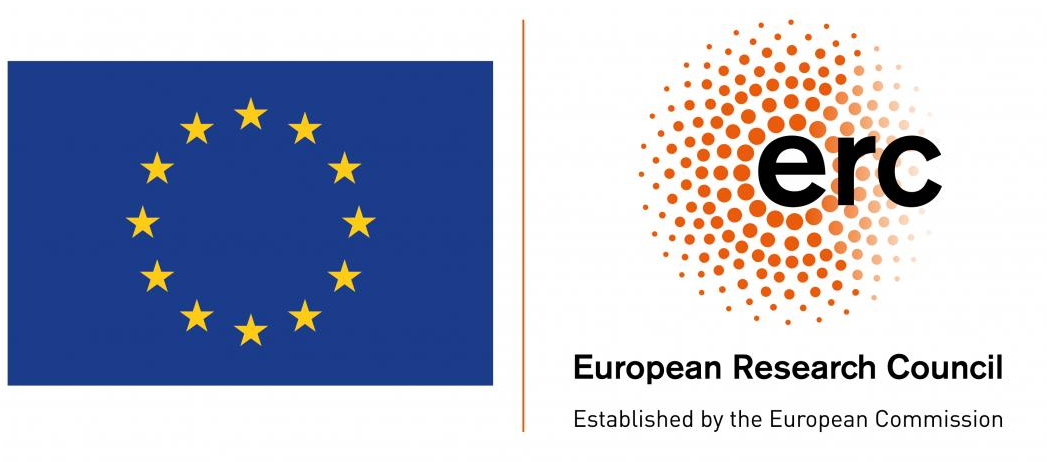}
			\end{figure}

			\bibliographystyle{abbrvnat}	\bibliography{aaai2027_biblio_cleaned}

@article{yang2024designing,
  author    = "Yang, Joshua C. and Hausladen, Carina I. and Peters, Dominik and Pournaras, Evangelos and Hnggli Fricker, Regula and Helbing, Dirk",
  title     = "Designing Digital Voting Systems for Citizens: Achieving Fairness and Legitimacy in Participatory Budgeting",
  journal   = "Digital Government: Research and Practice",
  volume    = 5,
  number    = 3,
  pages     = "1--30",
  year      = 2024,
  publisher = "ACM",
}

@inproceedings{masavrik2024generalised,
  author    = "Masarík, Tomáš and Pierczyński, Grzegorz and Skowron, Piotr",
  title     = "A Generalised Theory of Proportionality in Collective Decision Making",
  booktitle = "Proceedings of the ACM Conference on Economics and Computation",
  pages     = "734--754",
  year      = 2024,
}

@inproceedings{aziz2025committee,
  author    = "Aziz, Haris and Lederer, Patrick and Peters, Dominik and Peters, Jannik and Ritossa, Angus",
  title     = "Committee Monotonicity and Proportional Representation for Ranked Preferences",
  booktitle = "Proceedings of the ACM Conference on Economics and Computation",
  pages     = "896--896",
  year      = 2025,
}

@article{peters2025candidate,
  author    = "Peters, Jannik",
  title     = "Candidate Monotonicity and Proportionality for Lotteries and Non-Resolute Rules",
  journal   = "Social Choice and Welfare",
  pages     = "1--15",
  year      = 2025
}

@inproceedings{brill2023robust,
  author    = "Brill, Markus and Peters, Jannik",
  title     = "Robust and Verifiable Proportionality Axioms for Multiwinner Voting",
  booktitle = "Proceedings of the ACM Conference on Economics and Computation",
  pages     = "301--301",
  year      = 2023,
}

@article{aziz2020expanding,
  author    = "Aziz, Haris and Lee, Barton E.",
  title     = "The Expanding Approvals Rule: Improving Proportional Representation and Monotonicity",
  journal   = "Social Choice and Welfare",
  volume    = 54,
  number    = 1,
  pages     = "1--45",
  year      = 2020
}

@article{lederer2025proportional,
  author = "Lederer, Patrick",
  title  = "Proportional Representation in Rank Aggregation",
  journal = "arXiv preprint",
  year   = 2025,
  note   = "arXiv:2508.16177",
}

@inproceedings{aziz2021proportionally,
  author    = "Aziz, Haris and Lee, Barton E.",
  title     = "Proportionally Representative Participatory Budgeting with Ordinal Preferences",
  booktitle = "Proceedings of the AAAI Conference on Artificial Intelligence",
  pages     = "5110--5118",
  year      = 2021,
}

@article{halpern2026representation,
  author    = "Halpern, Daniel and Kehne, Gregory and Procaccia, Ariel D. and Tucker-Foltz, Jamie and Wüthrich, Manuel",
  title     = "Representation with Incomplete Votes",
  journal   = "Theory and Decision",
  pages     = "1--40",
  year      = 2026
}

@article{kellerhals2024proportional,
  author  = "Kellerhals, Leon and Peters, Jannik",
  title   = "Proportional Fairness in Clustering: A Social Choice Perspective",
  journal = "Advances in Neural Information Processing Systems",
  volume  = 37,
  pages   = "111299--111317",
  year    = 2024,
}

@inproceedings{papasotiropoulos2026fairness,
  author    = "Papasotiropoulos, Georgios and Pishbin, Zein",
  title     = "Fairness in the Multi-Secretary Problem",
  booktitle = "Proceedings of the AAAI Conference on Artificial Intelligence",
  pages     = "17188--17196",
  year      = 2026,
}

@inproceedings{mavrov2023fair,
  author    = "Mavrov, Ivan-Aleksandar and Munagala, Kamesh and Shen, Yiheng",
  title     = "Fair Multiwinner Elections with Allocation Constraints",
  booktitle = "Proceedings of the ACM Conference on Economics and Computation",
  pages     = "964--990",
  year      = 2023,
}

@inproceedings{Fain2018,
  author    = "Fain, Brandon and Munagala, Kamesh and Shah, Nisarg",
  title     = "Fair Allocation of Indivisible Public Goods",
  booktitle = "Proceedings of the ACM Conference on Economics and Computation",
  pages     = "575--592",
  year      = 2018,
}

@incollection{Lindahl1958,
  author    = "Lindahl, Erik",
  title     = "Just Taxation--A Positive Solution",
  booktitle = "Classics in the Theory of Public Finance",
  pages     = "168--176",
publisher = "Springer",
  year      = 1958
}

@article{Droop1881,
  author    = "Droop, Henry Richmond",
  title     = "On Methods of Electing Representatives",
  journal   = "Journal of the Statistical Society of London",
  volume    = 44,
  number    = 2,
  pages     = "141--202",
  year      = 1881,
  publisher = "JSTOR",
}

@article{skowron2026general,
  author = "Skowron, Piotr",
  title  = "A General Theory of Proportionality with Additive Utilities",
  journal = "arXiv preprint",
  year   = 2026,
  note   = "arXiv:2602.08504",
}

@book{dummett1984,
  author    = "Dummett, Michael",
  title     = "Voting Procedures",
  year      = 1984,
  publisher = "Oxford University Press",
}

@article{woodall1994properties,
  author  = "Woodall, Douglas",
  title   = "Properties of Preferential Election Rules",
  journal = "Voting Matters",
  pages   = "8--15",
  year    = 1994,
}

@article{tideman1995single,
  author    = "Tideman, Nicolaus",
  title     = "The Single Transferable Vote",
  journal   = "Journal of Economic Perspectives",
  volume    = 9,
  number    = 1,
  pages     = "27--38",
  year      = 1995,
  publisher = "American Economic Association",
}

@techreport{janson2016phragmen,
  author      = "Janson, Svante",
  title       = "Phragmén's and Thiele's Election Methods",
  year        = 2016,
  type        = "Technical Report",
}

@inproceedings{lederer2024squared,
author = {Lederer, Patrick and Peters, Dominik and W\k{a}s, Tomasz},
title = {The Squared Kemeny Rule for Averaging Rankings},
year = {2024},
booktitle = {Proceedings of the ACM Conference on Economics and Computation},
pages = {755}
}

@inproceedings{bardal2025proportional,
  author    = "Bardal, Tuva and Brill, Markus and McCune, David and Peters, Jannik",
  title     = "Proportional Representation in Practice: Quantifying Proportionality in Ordinal Elections",
  booktitle = "Proceedings of the AAAI Conference on Artificial Intelligence",
  pages     = "13581--13588",
  year      = 2025,
}

@book{lackner2023multi,
  author    = "Lackner, Martin and Skowron, Piotr",
  title     = "Multi-Winner Voting with Approval Preferences",
  year      = 2023,
  publisher = "Springer",
}

@misc{oskars,
  author = "{Academy of Motion Picture Arts and Sciences}",
  title  = "Special Rules for the Best Picture of the Year Award",
  year   = 2026,
 howpublished = {\url{https://www.oscars.org/sites/oscars/files/2026-05/99th%20Best%20Picture%20Rules.pdf
}},
  note   = "Accessed: 2026-07-23",
}

@misc{rankedwiki,
  author = "Wikipedia",
  title  = "Ranked Voting",
  year   = 2026,
    howpublished = {\url{https://en.wikipedia.org/wiki/Ranked_voting
}},
  note   = "Accessed: 2026-07-23",
}

@misc{pbwiki,
  author = "Wikipedia",
  title  = "List of Participatory Budgeting Votes",
  year   = 2026,
    howpublished = {\url{https://en.wikipedia.org/wiki/List_of_participatory_budgeting_votes
}},
  note   = "Accessed: 2026-07-23",
}

@article{SANCHEZFERNANDEZ2026104503,
  author    = "Sánchez-Fernández, Luis and Elkind, Edith and Lackner, Martin and Fernández García, Norberto and Fisteus, Jesús A. and Basanta Val, Pablo and Skowron, Piotr",
  title     = "Proportional Justified Representation",
  journal   = "Artificial Intelligence",
  volume    = 353,
  pages     = "104503",
  year      = 2026,
}

@article{aziz2017justified,
  author    = "Aziz, Haris and Brill, Markus and Conitzer, Vincent and Elkind, Edith and Freeman, Rupert and Walsh, Toby",
  title     = "Justified Representation in Approval-Based Committee Voting",
  journal   = "Social Choice and Welfare",
  volume    = 48,
  number    = 2,
  pages     = "461--485",
  year      = 2017
}

@inproceedings{kraiczy2026proportional,
  author    = "Kraiczy, Sonja and Papasotiropoulos, Georgios and Pierczyński, Grzegorz and Skowron, Piotr",
  title     = "Proportional Committee Elections with Positive and Negative Votes",
  booktitle = "Proceedings of the ACM Conference on Economics and Computation",
  year      = 2026,
  note      = "Forthcoming",
}

@article{PBsurvey,
  author = "Rey, Simon and Schmidt, Felicia and Maly, Jan",
  title  = "The (Computational) Social Choice Take on Indivisible Participatory Budgeting",
  journal = "arXiv preprint",
  year   = 2023,
  note   = "arXiv:2303.00621",
}

@inproceedings{papasotiropoulos2024method,
  author    = "Papasotiropoulos, Georgios and Pishbin, Seyedeh Zeinab and Skibski, Oskar and Skowron, Piotr and Wąs, Tomasz",
  title     = "Method of Equal Shares with Bounded Overspending",
  booktitle = "Proceedings of the ACM Conference on Economics and Computation",
  year      = 2025,
}

@inproceedings{BrillFJL17phragmen,
  author    = "Brill, Markus and Freeman, Rupert and Janson, Svante and Lackner, Martin",
  title     = "Phragmén's Voting Methods and Justified Representation",
  booktitle = "Proceedings of the AAAI Conference on Artificial Intelligence",
  pages     = "406--413",
  year      = 2017,
}

@inproceedings{peters2020proportionality,
  author    = "Peters, Dominik and Skowron, Piotr",
  title     = "Proportionality and the Limits of Welfarism",
  booktitle = "Proceedings of the ACM Conference on Economics and Computation",
  pages     = "793--794",
  year      = 2020,
}

@inproceedings{skowron2021proportionality,
  author    = "Skowron, Piotr",
  title     = "Proportionality Degree of Multiwinner Rules",
  booktitle = "Proceedings of the ACM Conference on Economics and Computation",
  pages     = "820--840",
  year      = 2021,
}

@article{mccune2024,
  author = "McCune, David",
  title  = "Monotonicity Anomalies in Scottish Local Government Elections",
  pages  = "69--101",
  journal   = "Social Choice and Welfare",
  year   = 2024,
}

@inproceedings{faliszewski2023pabulib,
  author    = "Faliszewski, Piotr and Flis, Jarosław and Peters, Dominik and Pierczyński, Grzegorz and Skowron, Piotr and Stolicki, Dariusz and Szufa, Stanisław and Talmon, Nimrod",
  title     = "Participatory Budgeting: Data, Tools and Analysis",
  booktitle = "Proceedings of the International Joint Conference on Artificial Intelligence",
  pages     = "2667--2674",
  year      = 2023}

@inproceedings{localStability,
  author    = "Haris Aziz and Edith Elkind and Piotr Faliszewski and Martin Lackner and Piotr Skowron",
  title     = "The Condorcet Principle for Multiwinner Elections: From Shortlisting to Proportionality",
  booktitle = "Proceedings of the International Joint Conference on Artificial Intelligence",
  pages     = "84--90",
  year      = 2017}

@inproceedings{cheng2019group,
    author = {Cheng, Y. and Jiang, Z. and Munagala, K. and Wang, K.},
    booktitle = {Proceedings of the 2019 ACM Conference on Economics and Computation},
    pages = {263--279},
    title = {Group Fairness in Committee Selection},
    year = {2019}
  }

@inproceedings{jiang2019approx,
    author = {Z. Jiang and K. Munagala and K. Wang},
    booktitle = {Proceedings of the 52nd Symposium on Theory of Computing (STOC-2020)},
    pages = {463--472},
    title = {Approximately Stable Committee Selection},
    year = {2020}
}

@inproceedings{pier-sko:core-stable,
  author       = {Grzegorz Pierczynski and
                  Piotr Skowron},
  title        = {Core-Stable Committees Under Restricted Domains},
  booktitle    = {International Conference on Web and Internet Economics},
  pages        = {311--329},
  year         = {2022},
}

@misc{dong2026jointly,
  title={Jointly Satisfying Pareto Optimality and Justified Representation is NP-Hard in Approval-Based Multiwinner Voting},
  author={Chris Dong},
  year={2026},
  eprint={2609.08357},
  archivePrefix={arXiv},
  primaryClass={cs.GT},
  doi={10.48550/arXiv.2609.08357},
  url={https://arxiv.org/abs/2609.08357}
}

			\appendix
			\onecolumn
			\section{Technical Appendix}

			\subsection{The Modified Solid Coalition Refinement Rule}\label{sec:modified_solid_coalition_rule}

			Both our theoretical results and empirical evaluation show that the Solid Coalition Refinement rule is weaker than the state-of-the-art proportional rules, in the case of approval ballots. A natural question is whether SCR can be modified to preserve its strong guarantees for ranked ballots while also satisfying stronger approval-based axioms.\looseness-1

			One natural approach is to change the criterion used to select a refined solid coalition. For $r \in \{1,\ldots,m\}$, let
			\begin{align*}
				\Top_i(r) = \{c \in C \colon |\{c' \in C \colon c' \succ_i c\}| < r\} \text{.}
			\end{align*}
			Thus, $\Top_i(r)$ contains the first $r$ positions in voter $i$'s ranking, including the entire indifference class at the boundary. Consequently, $|\Top_i(r)|$ may exceed $r$ when the ranking contains ties. Instead of using $\rho(W,S,X)$, the rule could select, at each step, a refined solid coalition maximizing the following quantity:
			\begin{align*}
				\widehat{\rho}(W,S,X)
				=
				\frac{|S|}
				{1+\max_{i\in S}|W\cap\Top_i(|X|)|}
				\text{.}
			\end{align*}
			This modification appears promising for two reasons:
			(i) For strict rankings, the modified rule coincides with the original SCR, so all our (axiomatic and algorithmic) results for strict rankings continue to apply.
			(ii) For approval committee elections, it coincides with the Greedy Justified Candidate Rule of~\cite{brill2023robust}, and therefore satisfies EJR+.

			Unfortunately, the construction below shows that this modification loses its proportionality guarantees in the presence of weak orders.
			In particular, it fails generalized PSC, even for the Hare quota, in the committee election setting, hence it also fails Proportionality (\Cref{def:prop}).
			Hence, while SCR performs sufficiently well for rankings, both strict and weak, but less well in the approval setting, the proposed modification performs well for both strict rankings and approval ballots, but not for weak rankings. Therefore, we view establishing a rule with strong guarantees across all three scenarios as the most important open question.

			Let $C=\{a,b,c,d,e\}$, and consider the committee elections setting with $k=4$ and $8$ voters.
			Consider the following preferences, where candidates in the same colored box belong to the same
			indifference class; colors indicate successive rank positions:
			\begin{center}
				\renewcommand{\arraystretch}{1.45}
				\begin{tabular}{c|l}
					voters & preferences\\
					\hline
					$1,3$
					&
					\rankone{a\sim b}
					$\succ$
					\ranktwo{c}
					$\succ$
					\rankthree{d}
					$\succ$
					\rankfour{e}
					\\
					$2$
					&
					\rankone{e}
					$\succ$
					\ranktwo{b}
					$\succ$
					\rankthree{a\sim c\sim d}
					\\
					$4$
					&
					\rankone{d}
					$\succ$
					\ranktwo{c\sim e}
					$\succ$
					\rankthree{a\sim b}
					\\
					$5$
					&
					\rankone{c}
					$\succ$
					\ranktwo{b}
					$\succ$
					\rankthree{a\sim d\sim e}
					\\
					$6$
					&
					\rankone{a\sim c}
					$\succ$
					\ranktwo{b}
					$\succ$
					\rankthree{d\sim e}
					\\
					$7$
					&
					\rankone{a\sim e}
					$\succ$
					\ranktwo{b}
					$\succ$
					\rankthree{c\sim d}
					\\
					$8$
					&
					\rankone{d}
					$\succ$
					\ranktwo{c}
					$\succ$
					\rankthree{e}
					$\succ$
					\rankfour{b}
					$\succ$
					\rankfive{a}
				\end{tabular}
			\end{center}

			Use a fixed tie-breaking order extending
			$bcde\succ_{\mathrm{tb}}abce\succ_{\mathrm{tb}}abcd
			\succ_{\mathrm{tb}}abc\succ_{\mathrm{tb}}bce
			\succ_{\mathrm{tb}}cde\succ_{\mathrm{tb}}cd
			\succ_{\mathrm{tb}}b\succ_{\mathrm{tb}}c
			\succ_{\mathrm{tb}}e\succ_{\mathrm{tb}}a
			\succ_{\mathrm{tb}}d$, where, for example,
			$bcde$ denotes the set $\{b,c,d,e\}$.

			A direct calculation gives the following refinement chains. At every arrow,
			the displayed pair indicates the set of voters $S$ in a solid coalition that maximizes $\hat{\rho}$ and
			the corresponding value of $\widehat{\rho}$:

			\begin{align*}
				W_0=\emptyset:\quad
				&C
				\xrightarrow{\substack{S^*=\{2,4,5,8\}\\ \widehat{\rho}=4}}
				bcde
				\xrightarrow{\substack{S^*=\{2,5\}\\ \widehat{\rho}=2}}
				bce
				\xrightarrow{\substack{S^*=\{1,3\}\\ \widehat{\rho}=2}}
				\{b\},
				\\[2mm]
				W_1=\{b\}:\quad
				&C
				\xrightarrow{\substack{S^*=\{2,4,5,8\}\\ \widehat{\rho}=2}}
				bcde
				\xrightarrow{\substack{S^*=\{4,8\}\\ \widehat{\rho}=2}}
				cde
				\xrightarrow{\substack{S^*=\{4,8\}\\ \widehat{\rho}=2}}
				cd
				\xrightarrow{\substack{S^*=\{5,6\}\\ \widehat{\rho}=2}}
				\{c\},
				\\[2mm]
				W_2=\{b,c\}:\quad
				&C
				\xrightarrow{\substack{S^*=\{2,7\}\\ \widehat{\rho}=2}}
				\{e\},
				\\[2mm]
				W_3=\{b,c,e\}:\quad
				&C
				\xrightarrow{\substack{S^*=\{1,3,6,7\}\\ \widehat{\rho}=2}}
				\{a\}.
			\end{align*}
			Hence the final outcome is $W=\{a,b,c,e\}$.
			Now consider $S=\{4,8\}$, $X=\{d\}$, and $\ell=1$. Both voters rank $d$
			uniquely first, so $(S,X)$ is a solid coalition that deserves one candidate according to both Droop and Hare quotas.
			Nevertheless,
			$|W\cap\periph_S(X)|
			=|\{a,b,c,e\}\cap\{d\}|=0<\ell$.
			Therefore the modified rule violates Proportionality.

			\subsection{Omitted Proofs} \label{sec:proofs}

			\subsubsection{Proof of \Cref{thm:drooppsc}.}

			Assume that $W$ satisfies generalized Droop-PSC and towards a contradiction say that $W$ violates Proportionality. Take the solid coalition $(S,X)$ that witnesses the violation. It holds that $(S,X)$ deserves $\ell$ candidates according to \Cref{def:prop} and $|W \cap \periph(S, X)| \leq \ell - 1$. Then $|W \setminus \periph(S, X)| \geq k - \ell + 1$ and we can pick $T \subseteq W \setminus \periph(S, X)$ with $|T| = k - \ell + 1$. Obviously, any $X' \subseteq X$ with $|X'| \geq \ell$ will give $|X' \cup T| \geq \ell + |T| = k + 1$. Therefore, the first condition of \Cref{def:prop} does not hold for $T$. As $(S,X)$ deserves $\ell$ candidates, it thus needs to hold that $\frac{|S|}{n} > \frac{\ell}{|T| + \ell} = \frac{\ell}{k+1}$, which is equivalent to $|S| > \ell \cdot \nicefrac{n}{k+1}$. With this, generalized Droop-PSC guarantees that $|W \cap \periph(S, X)| \geq \min(\ell, |X|)$. Taking $T = \emptyset$ in \Cref{def:prop} requires that $|X| \geq \ell$, which completes the contradiction.

			Assume that $W$ satisfies Proportionality and take a solid coalition $(S,X)$ such that $|S| > \ell \cdot \nicefrac{n}{k+1}$. We will first show that $(S,X)$ deserves $\ell' := \min(\ell, |X|)$ candidates. For this, consider any $T \subseteq W$. If $|T| + \ell' \leq k$, then we pick any $X' \subseteq X$ with $|X'| = \ell'$ towards satisfing \Cref{def:prop}. Otherwise $|T| + \ell' \geq k+1$, which allows us to satisfy \Cref{def:prop} with the inequality $\frac{|S|}{n} > \frac{\ell'}{k+1} \geq \frac{\ell'}{|T| + \ell'}$. These two cases complete the proof that $(S,X)$ deserves $\ell'$ candidates according to \Cref{def:prop}. Therefore, Proportionality guarantees $|W \cap \periph(S, X)| \geq \ell'$, as also required by generalized Droop-PSC.

			\subsubsection{Proof of \Cref{thm:pjr1}.}

			Assume that $W$ satisfies Proportionality and let $S$ be an arbitrary set of voters that deserve $\ell$ candidates according to PJR. Let $X=\bigcap_{i \in S} A_i$. Then $(S,X)$ is a solid coalition. Moreover, the requirement of deserving $\ell$ candidates according to PJR coincides with that under Proportionality. Hence, the fact that $W$ satisfies Proportionality implies that $|\periph(S,X)\cap W|\geq \ell$. By the choice of $X$, we have $\periph(S,X)=\bigcup_{i\in S}A_i$, and therefore, $W$ also satisfies PJR.

			\subsubsection{Proof of \Cref{thm:pjr2}.}

			Take an election $E=(C,N,\mathcal F, \mathcal A)$ such that $|W|=k$ for each inclusion-wise maximal $W \in \mathcal F$. Let $W$ be an inclusion-wise maximal feasible outcome satisfying \emph{PJR}. Towards showing that Proportionality is also satisfied, let $(S, X)$ be an arbitrary solid coalition deserving $\ell$ candidates in $W$ according to \Cref{def:prop}. We consider two cases. First, if $X \subseteq \bigcap_{i \in S} A_i$, then the group $S$ deserves $\ell$ candidates according to PJR, which guarantees $|W \cap \bigcup_{i \in S} A_i| \geq \ell$. But $\periph(S, X) = \bigcup_{i \in S} A_i$, so Proportionality is also satisfied. Otherwise, if $X \subsetneq \bigcap_{i \in S}$, then there exists a voter who has a non-approved candidate in $X$, which inflates the $\periph(S, X)$ to the whole set $C$. As $\ell \leq k$ is imposed by taking $T = \emptyset$ in the \Cref{def:prop}, we now have $|W \cap \periph(S, X)| = |W \cap C| = |W| = k \geq \ell$, again satisfying Proportionality.

			For the second part of the proof, assume that there exist two inclusion-wise maximal sets $A, B \in \mathcal F$ such that $|A| > |B|$. Consider an election with $|A|$ voters, each approving a different candidate in $A$. Let $W = B$. We will show that $W$ satisfies PJR but not Proportionality.

			According to PJR, a group of voters $S$ with $|S|>1$ does not deserve any candidate in $W$, as taking $T=\emptyset$ would require them to find a feasible subset of candidates of at least 1 candidate in $\bigcap_{i \in S} A_i = \emptyset$. A group $S$ consisting of a single voter who approves a candidate from $A \setminus B$ also does not deserve any candidate, as when picking $T = B$, by inclusion-wise maximality, there is no set that can be added to $T$ while preserving feasibility and also $\frac{|S|}{n} = \frac{1}{|A|} \leq \frac{1}{|T|+1}$. Lastly, a group $S$ consisting of a single voter that approves a candidate from $A \cap B$ satisfies $|\bigcup_{i\in S}A_i \cap W|=1$ while $S$ doesn't deserve more than one as for $T=\emptyset$ we have $\frac{1}{n}\leq 1$ and $|\bigcap_{i\in S}A_i|=1$ so one can't find a set $X'$ of size $2$ satisfying the corresponding requirement of PJR.
			Thus, $W$ satisfies PJR.

			We now move to Proportionality. The solid coalition $(N, X)$ deserves $|X|$ candidates in $W$. To show it, take an arbitrary $T \subseteq W$. If $T = \emptyset$, then the group can simply pick $X$ as the feasible set. Otherwise, we have $\frac{|S|}{n} = 1 > \frac{|X|}{|T|+|X|}$. However, it also holds that $|W \cap \periph(S, X)| = |W \cap C| = |W| = |Y| < |X|,$ violating Proportionality.

			\subsubsection{Proof of \Cref{thm:scr-electable-prop}.}

			We begin by assuming that each voter has an initial budget of $1$ and that candidates are being bought using voters' budget, each at a price of $\frac{n}{k+1}$, under a payment scheme to be defined.
			In particular, for any $t=1,\ldots,k$, we will show the existence of a payment scheme that buys $W_t$ while satisfying the following:
			\begin{enumerate}[leftmargin=2em,topsep=0pt,itemsep=0pt]
				\item[(i.)] no voter is ever left with a negative budget after a purchase, and
				\item[(ii.)] for every voter $i$ and every solid coalition $(S,X)$ with $i\in S$ and $c \in X \setminus W_{t-1}$ for which $\{c\} \cup W_{t-1} \in \mathcal F$ such that $\rho(W_t, S, X) > \frac{n}{k+1}$, then voter $i$ has spent money only on candidates in $\periph(S, X)$.
			\end{enumerate}

			Before proving the existence of such a payment scheme, we first show that it is sufficient to establish the statement of the lemma. In particular, we assume, towards a contradiction, that a payment scheme satisfying conditions (i) and (ii) exists, and that after step $k$ of the rule there is a solid coalition $(S, X)$ such that $\rho(W_k, S, X) > \frac{n}{k+1}$ and there exists $c \in X \setminus W_{k-1}$ with ${c} \cup W_{k-1} \in \mathcal{F}$. This means that
			\begin{align}
				\frac{n}{k+1} &< \frac{|S|}{|W_k \cap \periph(S, X)|+1} \notag\\
				n \cdot (|W_k \cap \periph(S, X)| + 1) &< |S| \cdot (k+1) \notag\\
				n \cdot (k+1) + n \cdot |W_k \cap \periph(S, X)| &< nk + |S|\notag \cdot (k+1) \\
				(n - |S|)(k+1) &< n \cdot (k - |W_k \cap \periph(S, X)|)\notag \\
				(n - |S|)(k+1) &< n \cdot |W_k \setminus \periph(S, X)|\notag \\
				n - |S| &< \frac{n}{k+1} \cdot |W_k \setminus \periph(S, X)|. \label{eq:n-s}
			\end{align}
			In the payment scheme for $W_k$ the voters from $S$ had only paid for the candidates from $\periph(S, X)$ (condition (ii)), which means that the voters from $N \setminus S$ had to pay for the whole $W_k \setminus \periph(S, X)$. Therefore they have to have paid at least $\frac{n}{k+1} \cdot |W_k \setminus \periph(S, X)|$ which, by condition (i), contradicts \Cref{eq:n-s}.

			We now turn to proving the existence of a payment scheme satisfying conditions (i) and (ii). The proof proceeds by induction on $t$.
			For $t=1$ we note that $W_0= \emptyset$, no purchase is made and no voter spends anything. Therefore, condition (i) holds trivially, and condition (ii) holds vacuously since no voter has spent any money.
			Assuming that a valid payment scheme exists for $W_{t-1}$, we show how to extend the payments to $W_t$ while still satisfying conditions (i) and (ii), by reasoning about how to pay for the inclusion of $c_t$ in $W_t$, where $c_t$ is the $t$-th candidate selected by the SCR rule.

			At a step $t,$ we call a candidate $c$ electable as long as $\{c\}\cup W_{t-1}\in \mathcal{F}$.
			Towards determining the payment scheme, consider first the case where there is no solid coalition $(S, X)$ with an electable candidate in $X$ and for which $\rho(W_{t-1}, S, X) > \frac{n}{k+1}$, i.e., the case where all solid coalitions are already represented in $W_{t-1}$ in proportion to their size. Then, we deduce the money for buying $c_t$ in an arbitrary way, while ensuring satisfaction of condition (i).
			This is possible since the voters’ total initial budget is sufficient to pay for $k$ candidates (and therefore also for $t$ candidates, as $t \leq k$), since $n > k \cdot \frac{n}{k+1}$.
			Moreover, it holds that $\rho(W_t, S, X) \leq \rho(W_{t-1}, S, X)\leq \frac{n}{k+1}$, for each solid coalition $(S,X)$ with an electable candidate, so condition (ii) holds again vacuously.

			Suppose now that there is a solid coalition $(S,X)$ with an electable candidate $c \in X$ and for which $\rho(W_{t-1}, S, X) > \frac{n}{k+1}$, and let $(S^*, X^*)$ be the last such during the execution of the while loop of the rule (Step 4 in \Cref{alg:scr}). This is well defined, as one needs to be picked during the first refinement.
			By condition (ii) applied to $W_{t-1}$, the voters from $S^*$ have only spent money on candidates from $\periph(S^*, X^*)$.
			Each such candidate costs $\frac{n}{k+1}$, so the voters from $S^*$ have spent at most $|W_{t-1}\cap\periph(S^*, X^*)| \cdot \frac{n}{k+1}$.
			Moreover, by definition, we have that $\frac{|S^*|}{|W_{t-1}\cap\periph(S^*, X^*)|+1} = \rho(W_{t-1}, S^*,X^*)$
			and we have assumed that $\rho(W_{t-1}, S^*, X^*) > \frac{n}{k+1}$. By rearranging we get that $|S^*|>(|W_{t-1}\cap\periph(S^*, X^*)|+1) \cdot \frac{n}{k+1}$. Therefore, the voters in $S^*$ have a remaining budget of at least $\frac{n}{k+1}$, so they can pay for a candidate $c$ in a way that satisfies condition (i). In the payment scheme for $W_t$ we split the cost $\frac{n}{k+1}$ for buying $c$ arbitrarily between the voters from $S^*$.
			At what follows we will show that this payment scheme for $W_t$ satisfies condition (ii) as well.

			Towards a contradiction, let $(S',X')$ denote a solid coalition with $c' \in X' \setminus W_{t-1}$ for which $\{c'\} \cup W_{t-1} \in \mathcal F$ such that $\rho(W_t, S', X') > \frac{n}{k+1}$ and assume that some voter $i \in S'$ spent money on a candidate that does not belong to $\periph(S', X')$.
			Since $\rho(W_{t-1}, S', X') \geq \rho(W_t, S', X') > \frac{n}{k+1}$, we infer from the induction hypothesis that voter $i$ has not spent money on candidates outside of $\periph(S', X')$ during the first $t-1$ steps of SCR rule. Hence, voter $i$ spent money on candidate $c_t$ and $c_t \notin \periph(S', X')$.
			According to the defined payment scheme, only voters from $S^*$ spent money on the candidate $c_t$, therefore $i\in S^*$ for the solid coalition $(S^*, X^*),$ meaning that $i$ weakly prefers each candidate in $X^*$ to any other candidate.
			Because $c_t \notin \periph(S', X')$ it follows that $X' \subsetneq X^*$. Note also that $X^*$ contains at least two electable candidates, namely $c'$ and $c_t$, so it was not the terminating coalition in the while loop. From the definition of $(S^*, X^*)$, we know that the next solid coalition in the while loop, say $(S^\diamond, X^\diamond)$ had $\rho(W_{t-1}, S^\diamond, X^\diamond) \leq \frac{n}{k+1}$. But then the SCR rule would have selected the solid coalition $(S', X')$ instead, as $\rho(W_{t-1}, S', X') > \frac{n}{k+1} \geq \rho(W_{t-1}, S^\diamond, X^\diamond)$, which contradicts the selection made by SCR, thereby proving the existence of the required payment scheme and concluding the proof.

			\subsubsection{Proof of \Cref{thm:lem2prop}.}

			We prove the result by induction on $k$. Recall that, according to \Cref{def:prop}, a solid coalition $(S, X)$ deserves $\ell$ candidates in $W$ if for every $T \subseteq W$ either there exists a set $X' \subseteq X$ with $|X'| \geq \ell$ such that $X' \cup T \in \mathcal F$ or the following inequality holds:
			\[ \frac{|S|}{n} > \frac{\ell}{|T| + \ell} \]
			For $k=0$, we have $W_0 = \emptyset$. Consider the set $T=\emptyset \subseteq W_0$. For this, the second condition becomes $\frac{|S|}{n} > 1$, which does not hold for any $S\subseteq N$.
			Therefore, any solid coalition $(S,X)$ that deserves $\ell$ candidates in $W$ must satisfy the first condition: there exists $X' \subseteq X$ with $|X'| \geq \ell$ and $X' \in \mathcal{F}$. Setting $X_0 = X'$ immediately gives the desideratum, since $X_0 \subseteq X \setminus W_0$ and $|W_0 \cap \periph(S, X)| = 0$.

			We now move to the case of $k \geq 1$ and we assume that the statement holds for $k-1$. Hence, there exists a set $X_{k-1} \subseteq X \setminus W_{k-1}$ such that $|X_{k-1}| + |W_{k-1} \cap \periph(S, X)| \geq \ell$ and $X_{k-1} \cup W_{k-1} \in \mathcal F$. Based on this, we need to show that there exists a set $X_k \subseteq X \setminus W_k$ such that $|X_k| + |W_k \cap \periph(S, X)| \geq \ell$ and $X_k \cup W_k \in \mathcal F$. Let $c_k$ be the candidate in $W_k\setminus W_{k-1}$.
			At what follows, we will make use of the following observation from \citet{masavrik2024generalised}.

			\begin{lemma}\label{take-one-lemma}
				For a matroid family $\mathcal F$, for all feasible sets $W \in \mathcal F$, $c \notin W$ and $W' \subseteq W$ ($W' \neq \emptyset$) such that $W \setminus W' \cup \{c\} \in \mathcal F$, there exists $c' \in W'$ such that $W \setminus \{c'\} \cup \{c\} \in \mathcal F$.
			\end{lemma}

			If $c_k \in \periph(S, X)$ then \Cref{take-one-lemma} says that there exists a candidate $c \in X_{k-1}$ such that $X_{k-1} \setminus \{c\} \cup W_k \in \mathcal F$. This means we can take $X_k = X_{k-1} \setminus \{c\}$.

			Now, say that $c_k \notin \periph(S, X)$. Then, if $|W_k \cap \periph(S, X)| \geq \ell$ we can simply use $X_k = \emptyset$. On the other hand, say that $|W_k \cap \periph(S, X)| + 1 \leq \ell$. Take $T = W_k \setminus \periph(S, X)$ and consider the answer of $S$ to $T$.
			We have that
			$\frac{|S|}{|W_k \cap \periph(S, X)|+1} \geq \frac{|S|}{\ell}.$
			The function $t \mapsto \frac{t}{|T| + t}$ is increasing, so:
			\begin{align*}
				\frac{\ell}{|T| + \ell}
				\geq \frac{|W_k \cap \periph(S, X)| + 1}{|T| + |W_k \cap \periph(S, X)| + 1}.
			\end{align*}
			By the definition of $T$, the denominator on the right satisfies
			$|T| + |W_k \cap \periph(S, X)| + 1 = |W_k| + 1 = k + 1$, giving:
			\begin{align}
				\frac{|S|}{n} > \frac{|W_k \cap \periph(S, X)| + 1}{k+1}
				\Leftrightarrow
				\rho(W_k, S, X) = \frac{|S|}{|W_k \cap \periph(S, X)| + 1} > \frac{n}{k+1}.
				\label{eq:largerho} \end{align}
			Since $|W_k \cap \periph(S, X)| + 1 \leq \ell$, the induction hypothesis $|W_{k-1} \cap \periph(S, X)| + |X_{k-1}| \geq \ell$, together with $c_k \notin \periph(S, X)$, implies that $|X_{k-1}| \geq 1$. Consequently, $(S,X)$ had an electable candidate. This means that we can apply \Cref{thm:scr-electable-prop} to get a contradiction with \Cref{eq:largerho}.
			In turn, it cannot be the case that $\frac{|S|}{\ell} > \frac{n - |S|}{|T|}$.
			Thus, as $S$ deserves $\ell$ candidates, there needs to exist $X' \subseteq X$ with $|X'| \geq \ell$ such that $X' \cup W_t \setminus \periph(S, X) \in \mathcal F$. From \Cref{def:matroids} we get that there exists the $X_k$.

			\subsubsection{Proof of \Cref{thm:non-matroid}.}

			Given family of constraints $\mathcal F$, as it is not a matroid, there exist two sets $X,Y \in \mathcal F$ such that $|X| < |Y|$ and $X \cup \{c\} \notin \mathcal F$ for all $c \in Y \setminus X.$ Say that $|Y|=\ell$.
			Consider the instance where a set $S$ of $\frac{\ell}{\ell+1} n + 1$ voters ranks all candidates in $X \cup Y$ tied and above the remaining candidates, while all other voters rank $X$ tied at the top and the rest below, also tied. This construction is identical to the one used in \cite{masavrik2024generalised} for showing that Phragmen doesn't satisfy PJR for non-matroid families of constraints, in the approval setting. Instead, we apply SCR to the instance.

			Let us first prove that, at every step $k$ of SCR, as long as there exists an electable candidate $c \in X$, that is, a candidate satisfying $\{c\} \cup W_{k-1} \in \mathcal{F}$, we have $W_k \subseteq X$. We prove this by induction on $k$.
			For $k=0,$ we have $W_k = \emptyset,$ so the claim holds trivially. For $k \geq 1$, assume that there exists a candidate $c \in X$ such that $\{c\} \cup W_{k-1} \in \mathcal F$ and $W_{k-1} \subseteq X$.
			We note that for a fixed set $C'\subseteq C,$ a solid coalition $(S',C')$ has strictly lower underrepresentation value than a solid coalition $(S'',C')$ with $S''\supset S'$. Moreover, there are two families of voter maximal solid coalitions:
			\begin{itemize}
				\item a coalition $(N, X')$ for any $X' \subseteq X$,
				\item a coalition $(S, Z)$ for any $Z \subseteq X \cup Y$ such that $Z \cap Y \neq \emptyset$.
			\end{itemize}
			To determine how SCR operates on the proposed instance we compute underrepresentation values as follows:
			\[ \rho(W_{k-1}, N, X') = \frac{n}{|W_{k-1} \cap \periph(N, X')|+1} = \frac{n}{k} > \frac{|S|}{k} = \frac{|S|}{|W_{k-1} \cap \periph(S, Z)|+1} = \rho(W_{k-1}, S, Z) \]
			Which means that during the first iteration of the while loop, the rule will pick a coalition $(N, X')$ for some $X' \subseteq X$, as Step 5 ensures that $X\subsetneq C$. Therefore, a candidate from $X$ will be elected in this round, which proves that $W_k \subseteq X$.
			Consequently, the outcome of SCR contains no candidate from $Y$. Indeed, after all candidates from $X$ have been included in the committee $W$, we have $W \cup \{c\} \notin \mathcal{F}$ for every candidate $c \in Y$, since $X \cup \{c\} \notin \mathcal{F}$ for all $c \in Y \setminus X$, by construction.

			We now show that the solid coalition $(S, X \cup Y)$ deserves $\ell$ candidates in the outcome. Consider any $T \subseteq W$. If $T = \emptyset$, then we can take $X' = Y$, and immediately obtain a set $X' \subseteq X \cup Y$ with $|X'| \geq \ell$ such that $X' \cup T \in \mathcal{F}$.
			Otherwise, if $|T|\geq1$, then we have the following:
			\[ \frac{|S|}{n} = \frac{\ell}{\ell + 1} + \frac{1}{n} > \frac{\ell}{\ell + 1} \geq \frac{\ell}{\ell + |T|}. \]
			Since $T \subseteq C$ was arbitrary, it proves that the coalition $(S, X \cup Y)$ deserves $\ell$ candidates in any outcome. Nevertheless, we have
			$|W \cap \periph(S, X \cup Y)| = |W \cap (X \cup Y)| = |X| < |Y| = \ell,$
			which shows that SCR fails Proportionality.

			\subsubsection{Proof of \Cref{thm:app}.}

			Assume that we have some fixed $X \subseteq C$, and we want to find $S \subseteq N$ such that $(S, X)$ is a solid coalition that maximizes the underrepresentation value. We define $N_X$ as the set of voters who have no voter from $C\setminus X$ ranked strictly above any candidate from $X$, i.e. they have $X$ on top.
			We first note that, in the approval voting setting, for any solid coalition $(S,X)$ the corresponding periphery is independent of $X$, namely either $\periph(S, X) = \bigcup_{i \in S} A_i$ or $\periph(S, X)=C.$
			Therefore, each refinement step for a set $D$ needs to compute underrepresentation values for two types of candidate sets: (i) singleton sets, which lead to immediate election and thus termination of the refinement process, and (ii) sets that exclude exactly one candidate from the current set---yielding in total only polynomially many possibilities.

			It remains to show that the corresponding maximization of the underrepresentation value (Step 6 of \Cref{alg:scr}) can also be carried out in polynomial time. We do so by reducing it to the problem of deciding whether there exists a solid coalition $(S, X)$ such that $\rho(W, S, X) \geq y$.
			Notice that there are only $O(nm)$ possible values of $y$, since the underrepresentation value is determined by a fraction whose numerator can take $O(n)$ values and whose denominator can take $O(m)$ values.
			We define an auxilitary function $f : 2^{N_X} \rightarrow \mathbb{Q}$ as
			$f(S) =
			\left| W \cap \periph(S, X) \right| - \frac{|S|}{y}$ and we note that
			\begin{align*}
				f(S) \leq -1 &\Leftrightarrow y| W \cap \periph(S, X)| - |S| \leq -y \Leftrightarrow\\ y(| W \cap \periph(S, X)| + 1) \leq |S| &\Leftrightarrow y\leq \frac{|S|}{| W \cap \periph(S, X)| + 1} = \rho(W, S, \{c\}).
			\end{align*}
			Therefore, it holds that $\rho(W, S, X) \geq y$ if and only if $f(S) \leq -1$. Moreover, the function $f$ is submodular.
			To show this, consider $S' \subsetneq S \subsetneq N$ and $i \in N \setminus S$, then
			\begin{align*}
				f(S' \cup \{i\}) - f(S')
				&= |W \cap \periph(S' \cup \{i\}, X)| - |W \cap \periph(S', X)| - \nicefrac{1}{y} \\
				&= |(W \cap \left(\periph(S', X) \cup \periph(i, X))\right)| - |W \cap \periph(S', X)| - \nicefrac{1}{y} \\
				&= |(W \cap \left(\periph(i, X) \setminus \periph(S', X))\right)| - \nicefrac{1}{y} \\
				&\geq |(W \cap \left(\periph(i, X) \setminus \periph(S, X))\right)| - \nicefrac{1}{y} \\
				&= f(S \cup \{i\}) - f(S).
			\end{align*}

			Overall, in each iteration of the rule, we determine $X$ either by setting $X=\{c\}$ or $X=D \setminus \{c\}$ for every available candidate $c$, and then aim to find a suitable set $S$; among the considered solid coalitions we pick the one maximizing the underrepresentation value. For this, we consider all possible values of $y$, or apply a binary search over them. For each considered value of $y$, we iterate over all candidate sets $X$ defined above and introduce a function $f$ to check whether there exists a set $S$ such that $f(S) \leq -1$. We then fix the solid coalition $(S,X)$ corresponding to a set $S$ that satisfies this condition for the maximum possible value of $y$.

			\subsubsection{Proof of \Cref{thm:boundedties}.}

			Given an election $E$, let $w$ denote the maximum number of tied candidates in any preference level of any voter. We first present an algorithm that is FPT with respect to $w$, and then refine it to obtain an algorithm that is FPT with respect to the maximum number of tied candidates in all but the last preference levels.

			To solve the problem of determining the solid coalition $(S,X)$ that maximizes the underrepresentation value (Step 6 of \Cref{alg:scr}), it is enough to consider the sets $X \subseteq C:$ that are on top of the ranking of at least one voter; these can be at most $nm2^w$.
			In line with the proof of \Cref{thm:app}, we will check whether or not there exists a solid coalition $(S, X)$ such that $\rho(W, S, X) \geq y$. Again, it is enough to iterate over the pre-computed list of possible sets $X \subseteq C$. We define $N_X$ as the set of voters who have no voter from $C\setminus X$ ranked strictly above any candidate from $X$, i.e. they have $X$ on top, and then introduce the auxilitary function $f : 2^{N_X} \rightarrow \mathbb{Q}$ as $f(S) = |W \cap \periph(S, X)| - \frac{|S|}{y}$.
			Again, $\rho(W, S, X) \geq y$ if and only if $f(S) \leq -1$ and $f$ is submodular.

			Now, let $w'$ denote the maximum number of tied candidates in any non-last preference level of any voter. We can then solve the problem of determining the solid coalition $(S,X)$ that maximizes the underrepresentation value by considering two families of sets $X \subseteq C$: (i) those appearing at the top of the ranking of at least one voter, without including the bottom preference level, and (ii) those containing at least one candidate from the bottom preference level of some voter. For the first case, we proceed exactly as in the parameterization with respect to $w$. In the second case, we have $\periph(S,X)=C$, making it analogous to the case considered in the proof of \Cref{thm:app}, and introducing only an additional number of $m$ sets to consider.

			\subsubsection{Proof of \Cref{thm:FPTsmallelections}.}
			We denote by $\succ_i$ the strict preference relation between alternatives in the order submitted by voter $i$.
			For the first part of the theorem, we construct the family of all solid coalitions that need to be considered towards solving the maximization problem in Step 6 of \Cref{alg:scr}. To do so, we first iterate over all possible pairs $(X, X')$ such that $X \subseteq X' \subseteq C$; the set $X'$ serves as a possible (inclusion-wise) upper bound for the periphery. Having fixed the sets $X$ and $X'$, we compute a set of voters $S$ such that
			\( S = \{ i \in N \ :\ X \succeq_i C \setminus X \ \wedge X' \succ_i C \setminus X' \} \). Observe that $S$ is the maximal set of voters such that $(S, X)$ is a solid coalition satisfying $\periph(S, X) \subseteq X'$, and, for the specified pair $(X,X')$ only a single solid coalition, namely $(S,X),$ needs to be examined. In turn, the maximization function of the underrepresentation value needs to be employed over a family of size bounded by $3^m$.

			We now move to the second part of the theorem for which we will prove that the maximization problem in \Cref{alg:scr} can be solved in time $\mathcal O(2^n \cdot nm^2)$.
			We first observe that for a group of voters $S \subseteq N$ and $X, X' \subseteq C$ such that $|X| = |X'|$ and both $(S, X)$ and $(S, X')$ are solid coalitions, it holds that \( \periph(S, X) = \periph(S, X') \). This comes from the fact, that we can express the periphery solely in terms of $S$ and $r = |X|$. In particular,
			$\periph(S, X) = \bigcup_{i \in S} \{ c \in C \ :\ c \succeq_i c^{i,r} \},$
			where $c^{i,r}$ is the $r$-th candidate on voter $i$ ranking, breaking ties arbitrarily.

			We will show that can solve the maximization problem by first iterating through all $S \subseteq N$ and then by all possible sizes of $X$.
			For each such $S$ and a possible size $r$ we present a procedure to determine if there exists a set of candidates $X$ such that
			\begin{enumerate}[label=(\roman*),leftmargin=2em,topsep=0pt,itemsep=0pt]
				\item\label{itm:r} $|X| = r$,
				\item\label{itm:solid} $(S, X)$ is a solid coalition,
				\item\label{itm:X-sub-D} $X \subsetneq D$,
				\item\label{itm:X-sub-W} $X \cap C' \neq \emptyset$.
			\end{enumerate}
			Due to the previous observation, if multiple sets exist that satisfy these conditions, all of which will result in the same underrepresentation value. Therefore, it is sufficient to construct only one of them. If two candidates $c,c'$ belong to the same indifference class for a voter $i$ we write $c \sim_i c'$.
			Towards constructing a suitable set $X$, we define the following sets of candidates:
			\begin{equation*}
				X_{always} = \bigcup_{i \in S} \{ c \in C \ :\ c \succ_i c^{i,r} \}
				,\quad
				X_{maybe} = \bigcap_{i \in S}  \{ c \in C \ :\ c \sim_i c^{i,r}\}.
			\end{equation*}

			All candidates from the set $X_{always}$ will have to be included in the constructed set $X$. To argue that this is the case, focus on some voter $i \in S$. If she is a part of a solid coalition $(S, X)$ with $|X| = r$, then it needs to be the case that $\{ c \in C : c \succ_i c^{i,r} \} \subseteq X$, as the set $X$ has to be on top of her ranking, and to guarantee that $X$ satisfies condition~\ref{itm:solid}, we need to check that the set $X_{always}$ is on top of the ranking of every voter from $S$. Similarly, candidates from $C\setminus (X_{always}\cup X_{maybe})$ should not be included in $X$.
			The condition~\ref{itm:r} additionally requires that $|X_{always}| \leq r$.
			Further, the condition~\ref{itm:X-sub-D} requires that $X_{always} \subset D.$

			After determining that $X_{always} \subseteq X$, we need to fill the remaining $r - |X_{always}|$ spots. We do that, by adding some candidates from the set $X_{maybe} \cap D$. We prioritize electable candidates, towards satisfying condition~\ref{itm:X-sub-W}.
			At the end, we again verify if condition~\ref{itm:X-sub-W} holds.

			If any of the above checks fails, there cannot exist a set $X$ that satisfies all the introduced constraints. If the construction was successfull, we add the solid coalition $(S, X)$ to a list of coalitions for consideration. This scheme allows us to construct necessary coalitions for the maximization problem arising at \Cref{alg:scr}. The complexity of the algorithm with this subprocedure is $\mathcal{O}(2^n \cdot nm^4)$. This comes from the $\mathcal{O}(2^n \cdot m)$ iterations in the outer loops, an additional $\mathcal{O}(nm)$ cost for determining the sets $X_{\text{always}}$ and $X_{\text{maybe}}$, and up to $\mathcal{O}(m)$ refinements for each of the $\mathcal{O}(m)$ candidates to be potentially elected. Overall, the running time is FPT with respect to the number of voters.

			\subsubsection{Proof of \Cref{thm:propdegree}.}

			Let $t = \lfloor\sqrt k\rfloor$ and $\ell=t-1$. Consider an election with $C = \{ c_1, \ldots, c_k, d_1, \ldots, d_\ell \}$ and $N = \{1, \ldots, k\}$. We introduce two groups of voters, with approval ballots defined as follows:
			\[ A_i = \begin{cases}
				\{ c_i, d_1, \ldots, d_\ell \} & \text{if } i \leq \ell \\
				\{ c_1, \ldots, c_k \} & \text{if } i > \ell
			\end{cases} \]
			We denote the set of voters of the first group as $S_1$ and the set of voters of the second as $S_2$.

			We will show that the SCR rule elects ${c_1, \ldots, c_k}$ as the winning committee. Let $i \geq 0$ and assume that the candidates ${c_1, \ldots, c_i}$ have already been elected. Taking ties into account, there are two candidates under consideration for election, $c_{i+1}$ and $d_1$. We now consider two cases.

			If $i < \ell$, then the maximum underrepresentation value that can be achieved with respect to the candidate $c_{i+1}$ is \( \rho(W_i, S_2 \cup \{i + 1\}, \{c_{i+1}\}) = \frac{k - \ell + 1}{i + 1}\), to be denoted for simplicity by $\rho(c_{i+1})$. Note that this value is strictly greater than $\rho(W_i, \{i+1\}, \{c_{i+1}\}) = 1$. For the underrepresentaion value with respect to the candidate $d_1$, to be denoted for simplicity by $\rho(d_1)$,  consider the group of voters with $j \leq i$ of them approving some candidate that already have been elected. Then the underrepresentation value is
			$\rho(d_1) = \rho(W_i, \{1, 2, \ldots, j, i+1, i+2, \ldots, \ell \}, \{d_1\}) = \frac{\ell - i + j}{j + 1} $,
			which is maximal for $j = 0$.

			We want to show that
			$ \rho(d_1) = \ell - i < \frac{k - \ell + 1}{i + 1} = \rho(c_{i+1}).$
			Equivalently, we want to show that $-i^2+(\ell - 1)i + \ell < k-\ell+1.$

			We first claim that $\frac{(\ell+1)^2}{4} < k-\ell+1 \Leftrightarrow t^2 < 4(k-t+2)$.
			This is true because $3t^2-4t+8 = 3(t-\nicefrac{2}{3})^2+\nicefrac{20}{3}>0 \Leftrightarrow t^2+4m-8<4t^2\leq 4k,$
			where the last inequality follows from $k\geq t^2$.

			Consider the function $f(i)=-i^2+(\ell - 1)i + \ell,$ for $i \in [0,\ell)$. It holds that $f'(i)=-2i+\ell-1, f''(i)<0$. Hence, $f$ has its maximum at $i^*=\nicefrac{\ell-1}{2}$.
			Moreover $f(i^*)=\nicefrac{(\ell+1)^2}{4}.$
			Therefore, $-i^2+(\ell - 1)i + \ell \leq \nicefrac{(\ell+1)^2}{4}<k-\ell+1$.

			For the second case of $i \geq \ell$, every voter from $S_1$ already got a satisfaction with respect to one candidate, which leads to $\rho(W_i, S, {d_1}) = \frac{|S|}{|S|+1}$ for any $S \subseteq S_1$; we denote it by $\rho(d_1)$. Regarding $c_{i+1}$, we also compute the underrepresentation value, to be denoted by $\rho(c_{i+1})$ and for this we note that only voters from $S_2$ approve $c_{i+1}$. The highest underrepresentation values for these candidates are achieved by groups $S_1$ and $S_2$, respectively. Therefore, we need to show that
			\[ \rho(d_1) = \frac{\ell}{\ell + 1} < \frac{k - \ell}{i+1} = \rho(c_{i+1}).\]
			First, we claim that $k-\ell > \ell$. This holds because $t^2\leq k$ and $t^2 >2t-2,$ so $k\geq 2t-2 \Leftrightarrow k-t+1>t-1.$
			We want to prove that $\frac{\ell}{\ell + 1} < \frac{k - \ell}{i+1},$ or, equivalently, that $2\ell^2+\ell < k\ell+k.$
			Given that $k-\ell > \ell,$ we have $k(\ell+1)>2\ell(\ell+1)>2\ell^2+\ell.$

			Consequently, in both cases, the candidate $c_{i+1}$ will be elected.
			We also have to check the underrepresentation value arising from the case where a group of voters $S$ ranks a candidate $c \in C$ at the bottom of their ranking, i.e., disapproves of $c$, and may therefore contribute to refining the examined set of candidates $D$ to $D \setminus \{c\}$. In this case, $S$ is necessarily a subset of either $S_1$ or $S_2$, and therefore the corresponding underrepresentation ratio is no larger than the ones already computed.

			By induction, we get that the SCR rule will select the committee $\{c_1, \ldots, c_k\}$. To conclude the proof, observe that $|\bigcap_{i \in S_1} A_i| = |\{d_1, \ldots, d_\ell\}| = \ell$, $|S_1| = \ell = \ell \cdot \frac{n}{k}$ and for each $i \in S_1$ we have $|A_i \cap SCR(E)| = 1 = \ell \cdot \frac{1}{\lfloor\sqrt{k}\rfloor - 1}$

			\subsubsection{Proof of \Cref{thm:propdegreecomp}.}

			In the construction of \Cref{thm:propdegree}, each voter in the group $S_1$ approves a different candidate $c_i$, and $|S_1|=\ell$. Moreover, each voter in $S_1$ obtains a satisfaction of $1$ from the outcome of SCR, while deserving $\ell$. Therefore, if the size of $S_1$ could be increased while keeping the satisfaction of its voters bounded by one, the resulting proportionality degree would decrease.
			To this end, let $S_1$ be partitioned into $k$ groups such that the $i$-th group has size equal to $x_i \cdot n$ and its members approve candidate $c_i$, as well as enough candidates from the set $\{d_1, d_2, \ldots \}$ so that the cohesiveness of $S_1$ can only be limited by its size.
			We fix the committee size $k$ and formulate the problem of maximizing the size of $S_1$—which in turn yields an upper bound on the proportionality degree of SCR—as a linear program.
			\begin{align*}
				\text{maximize:}\quad
				& \sum_{i=1}^k x_i \\
				\text{with respect to:}\quad
				& \sum_{i=1}^k x_i \leq 1 \\
				& x_i \geq 0 && \forall i \in \{1, \ldots, k\} \\
				& x_{i+1} \leq x_{i} && \forall i \in \{1, \ldots, k-1\} \\
				& \frac{(1-\sum_{i=1}^{k}x_i) + x_t}{t} \geq \frac{\sum_{j=1}^r x_i + \sum_{i=t}^k x_i}{r+1}
				&& \forall t \in \{1, \ldots, k\}, \forall r \in \{0, \ldots, t\}
			\end{align*}
			Here, the variables represent the fraction of voters in $S_1$ relative to the entire electorate, so that $\sum_{i=1}^k x_i = \frac{|S_1|}{n}$.
			We have to ensure that the candidate elected at step $t$ is $c_t$, which is guaranteed by the last inequality. The left-hand side captures the underrepresentation value with respect to candidate $c_t$, who is supported by all voters in $S_2$ as well as by some voters from $S_1$ which amount to an $x_t$ fraction of the voters of the electorate. The right-hand side captures the underrepresentation value with respect to candidate $d_1$, who is supported by all voters in $S_1$ that have not yet received a representative, together with those groups whose representative has already been selected.
			The third constraint ensures that the candidates are selected in order of increasing index. In particular, the first selected candidate is the one supported by the largest fraction of voters from $S_1$, the second selected candidate is the one supported by the second-largest fraction, and so on.
			Solving the program for small values of $k$ already suggests a substantially stronger upper bound that is less than $\frac{3}{k^{0.72}}$, as the following plot demonstrates.

			\begin{center}
				\includegraphics[width=0.52\textwidth]{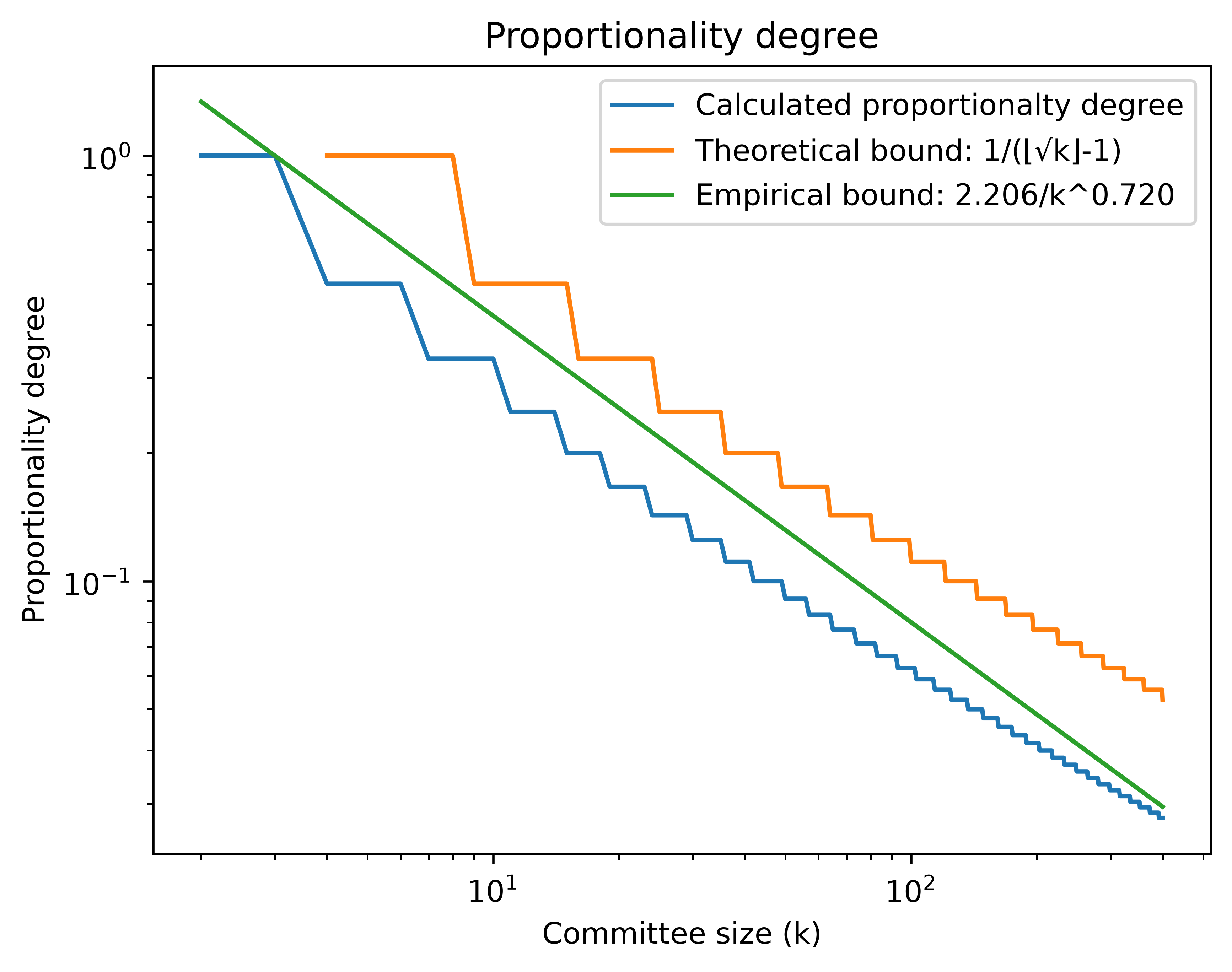}
			\end{center}

			\subsection{Additional Findings and Details} \label{sec:app_exp}

We start by noting that SCR might produce outcomes that are undesirable from a basic efficiency perspective, even in simple instances, specifically, outcomes that are not Pareto optimal \cite{lackner2023multi}.

\begin{definition}
	Let $E=(C, N, \{ W' \subseteq C : |W'| \leq k \}, \mathcal A)$ be a committee election under approval ballots. An outcome $W$ is Pareto dominated by an outcome $W'$
	if for each voter $i \in N$ it holds that
	$|A_i\cap W| \leq |A_i\cap W'|$ while for at least one
	voter the inequality is strict. An outcome $W$ is Pareto optimal if it is
	not Pareto dominated by any feasible outcome.
\end{definition}

This result is not surprising, especially in light of the recent result by \citet{dong2026jointly}. Since SCR is polynomial-time computable and satisfies PJR in approval committee elections, it cannot also satisfy Pareto optimality unless P=NP. Our result, however, is unconditional. Moreover, the same instance illustrates the difference between the sequential decisions of seqPAV and SCR. While both iteratively select candidates based on the representation achieved by previous selections, seqPAV tracks individual voters’ satisfaction, whereas SCR evaluates solid coalitions through their collective periphery. Consequently, a voter may be considered represented through their coalition even if they do not individually enjoy the same satisfaction, and this explains how SCR can fail Pareto optimality.

\begin{theorem}
	The outcome of the SCR rule is not always Pareto optimal, even for approval committee elections.
	\label{thm:PO}
\end{theorem}
	\begin{proof}
		Consider an approval committee election with
		$C=\{a,b,c,d\}$, $k=2$, and the following $14$ voters:
		\begin{center}
			\begin{tabular}{c|c|c}
				Group & Number of voters & Approval set \\
				\hline
				$G_1$ & $5$ & $\{a,c\}$ \\
				$G_2$ & $3$ & $\{a,d\}$ \\
				$G_3$ & $3$ & $\{b,d\}$ \\
				$G_4$ & $2$ & $\{b,c\}$ \\
				$G_5$ & $1$ & $\{d\}$
			\end{tabular}
		\end{center}
		We show that SCR returns $W=\{a,b\}$ which is Pareto dominated by $W'=\{c,d\}$.
		Recall that, under approval
		preferences, a voter $i$ solidly supports a nonempty set $X$ if and only if
		$X\subseteq A_i$ or $A_i\subseteq X$. Moreover, if every voter in a solid
		coalition $(S,X)$ approves every candidate in $X$, then
		$
		\periph_S(X)=\bigcup_{i\in S}A_i.
		$
		Otherwise, $\periph_S(X)=C$.
		
The following table gives the maximum value of
		$\rho(W,S,X)$ over all nonempty solid coalitions $(S,X)$, for each
		nonempty proper subset $X$ of $C$, at the two partial outcomes that
		occur in the execution. A dash means that no nonempty group solidly
		supports $X$.
		\begin{center}
			\begin{tabular}{c|c|c}
				$X$
				& $\max_S \rho(\emptyset,S,X)$
				& $\max_S \rho(\{a\},S,X)$ \\
				\hline
				$\{a\}$     & $8$   & $(5+3)/2=4$ \\
				$\{b\}$     & $5$   & $3+2=5$ \\
				$\{c\}$     & $7$   & $\max\{2,(5+2)/2\}=7/2$ \\
				$\{d\}$     & $7$ & $\max\{3+1,(3+3+1)/2\}=4$ \\
				\hline
				$\{a,b\}$   & --        & -- \\
				$\{a,c\}$   & $5$       & $5/2$ \\
				$\{a,d\}$   & $4$   & $(3+1)/2=2$ \\
				$\{b,c\}$   & $2$       & $2$ \\
				$\{b,d\}$   & $4$   & $\max\{3,(3+1)/2\}=3$ \\
				$\{c,d\}$   & $1$       & $1/2$ \\
				\hline
				$\{a,b,c\}$ & $7$   & $(5+2)/2=7/2$ \\
				$\{a,b,d\}$ & $7$ & $(3+3+1)/2=7/2$ \\
				$\{a,c,d\}$ & $9$ & $(5+3+1)/2=9/2$ \\
				$\{b,c,d\}$ & $6$ & $(3+2+1)/2=3$
			\end{tabular}
		\end{center}
		
		Initially, $W_0=\emptyset$ and $D=C$. Since
		$\rho(\emptyset,S,X)=|S|$, the unique maximizing refinement is
		$
		(S,X)=(G_1\cup G_2\cup G_5,\{a,c,d\}),
		$
. Therefore, SCR first replaces $D$ by
		$\{a,c,d\}$. Notice that $(N,C)$ is not eligible, since a refinement
		must be a proper subset of the current set $D$.
				The nonempty proper subsets of the new $D$ are
		$
		\{a\}, \{c\}, \{d\}, \{a,c\}, \{a,d\},$ $\{c,d\}.
		$
		Thus, the unique maximizing refinement is
		$(G_1\cup G_2,\{a\})$, and SCR elects $a$.
		
		Consider now the second round, where $W_1=\{a\}$ and $D=C$.
		As shown in the table, 
		the unique maximizing refinement is
		$(G_3\cup G_4,\{b\})$. Therefore, SCR elects $b$ and returns
		$
		W=\{a,b\}.
		$

	We conclude by noting that, on the same instance, sequential PAV also uniquely elects $a$ first.
	After electing $a$, the marginal scores are
	\[
	\Delta(b\mid\{a\})=3+2=5,\qquad
	\Delta(c\mid\{a\})=\frac52+2=\frac92,\qquad
	\Delta(d\mid\{a\})=\frac32+3+1=\frac{11}{2}.
	\]
	Therefore, sequential PAV uniquely elects $d$ next and returns
	$\{a,d\}$, rather than the SCR outcome $\{a,b\}$. The difference is
	that sequential PAV adds the half-weight contributions of the three
	already represented voters in $G_2$ to the four unrepresented
	supporters of $d$, whereas SCR evaluates these supporters through
	a common coalition-level denominator. Notice also that $\{a,d\}$
	is Pareto optimal on this instance: any committee Pareto dominating
	it would have to give the voters in $G_2$ two approved winners,
	and hence would have to contain both $a$ and $d$.
\end{proof}

		The remaining of this section is devoted to empirical findings and details.	In \Cref{sec:experiments} we analyzed the $k$-Borda and the Utilitarian Greedy algorithms which select the outcome $$W = \arg\max\limits_{W \in \mathcal F} \sum\limits_{i \in N} u_i(W),$$ where $u_i(W) = \sum_{c \in W} \left(m - \rank_i(c)\right)$ and $u_i(W) = |W \cap A_i|$ respectively. For $k$-Borda, we assumed that voters get a utility of $0$ from their unranked candidates, i.e., those positioned last, and $\rank_i(c)$ indicates the position of $c$ in the ranking of $i$.

			The Method of Equal Shares (MES) proceeds through a sequence of rounds. In approval-based committee elections, each voter $i$ is assigned a virtual budget $b_i$, initialized to $\nicefrac{k}{n}$, and every candidate is assumed to have unit cost.
			In each round, we consider the candidates that have not yet been elected and whose approving voters collectively have enough remaining budget to cover the cost of the candidate.
			More formally, for a candidate $c$ and a parameter $\rho \in \mathbb{R}_{+}$, we say that $c$ is \textit{$\rho$-affordable} if
			\[
			\sum_{i \in A_c} \min\left(b_i,\rho\right)=1.
			\]
			Among all candidates that are $\rho$-affordable for some value of $\rho$, the one with the smallest such $\rho$ is selected and added to the committee.
			The virtual budgets of its supporters are then updated:
			$b_i := b_i-\min\left(b_i,\rho\right).
			$
			The process terminates when there is no $\rho$-affordable candidate for any value of $\rho$.

			Sequential PAV (seqPAV) is a polynomial time computable variant of Proportional Approval Voting that constructs the committee sequentially. In each iteration, it selects the candidate that maximizes the marginal increase in voters' total satisfaction, where a voter's satisfaction from receiving multiple approved candidates follows a harmonic sequence.
			We also analyzed EAR and STV for the setting of ranked ballots. We present the corresponding pseudocodes as \Cref{ear} and \Cref{stv}.
			\begin{figure}[h!]
				\centering
				\begin{minipage}{0.48\linewidth}
					\centering
					\begin{algorithm}[H]
						\caption{The pseudocode of the Expanding Approvals Rule.}
						\label{ear}
						$W \gets \emptyset$

						$\{ w_i \gets 1 \}_{i \in N}$

						\For{$r = 1, \ldots, m$} {

							$\{ N_c \gets \{ i \in N : \rank_i(c) \leq r \} \}_{c \in C'}$

							$\{ r_c \gets \sum\limits_{i \in N_c} w_i \}_{c \in C'}$

							\While{$\max\limits_{c \in C \setminus W} r_c \geq q$} {
								$c' \gets \arg\max\limits_{c \in C'} r_c$

								\For{$i \in N_{c'}$} {
									$w_i \gets w_i \cdot (1 - \nicefrac{q}{r_{c'}})$
								}

								$W \gets W \cup \{c'\}$
							}
						}

						\Return $W$
					\end{algorithm}
				\end{minipage}
				\hfill
				\begin{minipage}{0.48\linewidth}
					\centering
					\begin{algorithm}[H]
						\caption{The pseudocode of the Single Transferable Vote rule.}
						\label{stv}
						$W \gets \emptyset$

						$C' \gets C$

						$\{ w_i \gets 1 \}_{i \in N}$

						\While{$C'\neq \emptyset$}{

							$\{ N_c \gets \{ i \in N : c \succ C' \setminus \{c\} \} \}_{c \in C'}$

							$\{ r_c \gets \sum\limits_{i \in N_c} w_i \}_{c \in C'}$

							\If{$\max\limits_{c \in C'} r_c \geq q$} {

								$c' \gets \arg\max\limits_{c \in C'} r_c$

								\For{$i \in N_{c'}$} {
									$w_i \gets w_i \cdot (1 - \nicefrac{1}{r_{c'}})$
								}

								$W \gets W \cup \{c'\}$

								$C' \gets C' \setminus \{c'\}$

							}

							\Else {
								$c' \gets \arg\min\limits_{c \in C'} r_c$

								$C' \gets C' \setminus \{c'\}$
							}
						}

						\Return $W$
					\end{algorithm}
				\end{minipage}
			\end{figure}

			We note that both EAR and STV have nuances on the quota used and how the cost is divided between the voters. As we said in the main text, we use Add1 completion, meaning that we iteratively decrease the quota as long as the rule returns a feasible outcome, trying to find minimal quota for which the rule returns a feasible committee.
			More precisely, in iteration $i$ we use quota $q_i = \frac{n}{k+0.01i}$. We stop at the first iteration the rule returns a committee of size greater than $k$ and use the committee from the previous iteration.
			We also considered the completion method that slightly increases voters’ utilities for non-approved projects (the so-called Eps method by \citet{faliszewski2023pabulib}), but it resulted in inferior performance for these rules.
			In both of the rules, when a candidate is elected, the quota needs to be paid by its supporters. We follow the recommendation of \citet{aziz2020expanding} of using \emph{uniform fractional reweighting}.

			The instances from \textsc{Pabulib} are partitioned into four groups of approximately equal size. In particular, our analysis covers the 410 instances from the library that satisfy $m>k$ and $k>1$ after the transformation to committee elections described above. For the Scottish local elections dataset, we note that a value of $k$ is provided for each instance, and we use these values in our experiments.

\begin{figure}[h!]
				\centering
				\begin{minipage}{0.48\textwidth}
					\centering
					\includegraphics[width=0.95\textwidth]{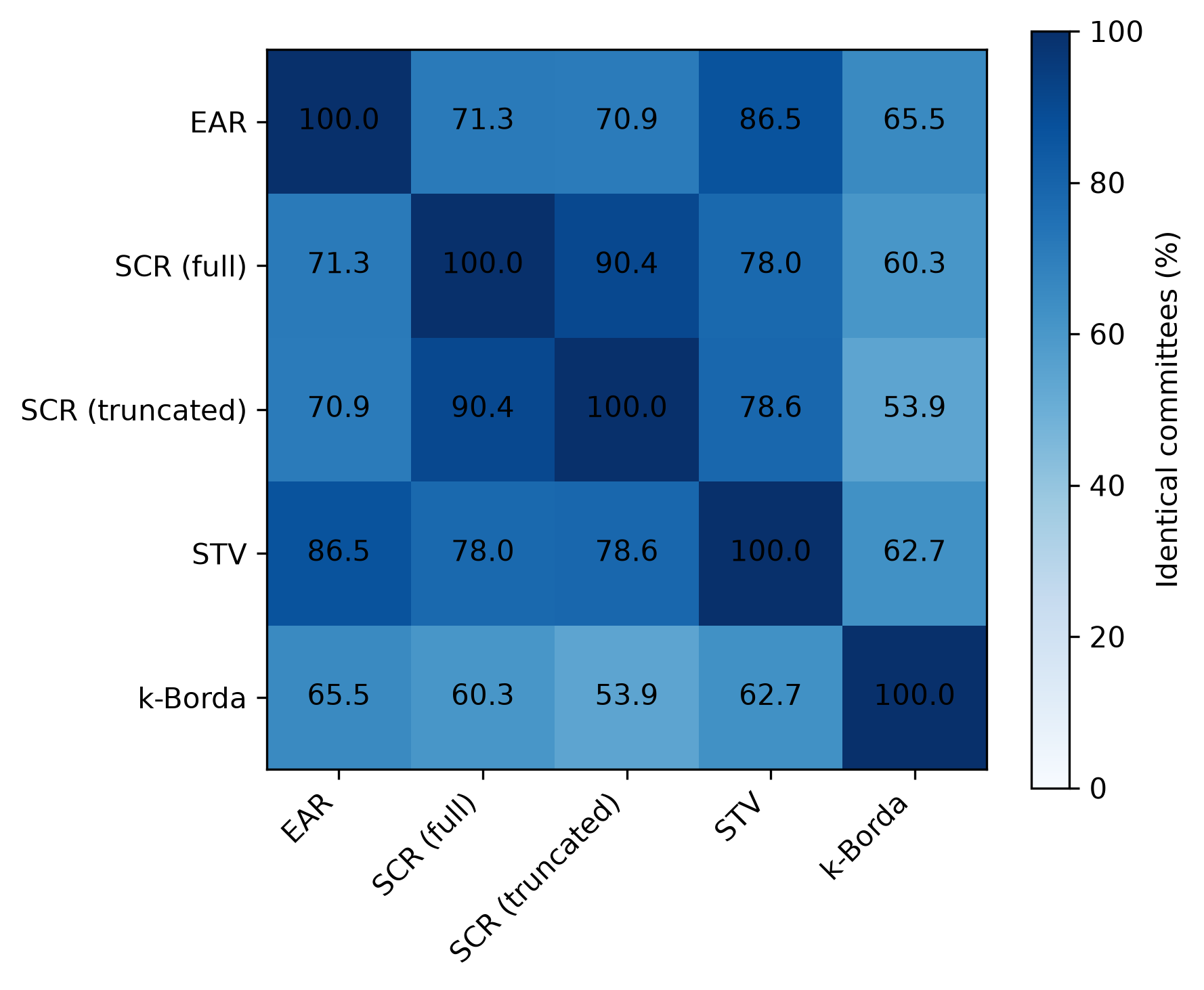}
				\end{minipage}
				\hfill
				\begin{minipage}{0.48\textwidth}
					\centering
					\includegraphics[width=0.95\textwidth]{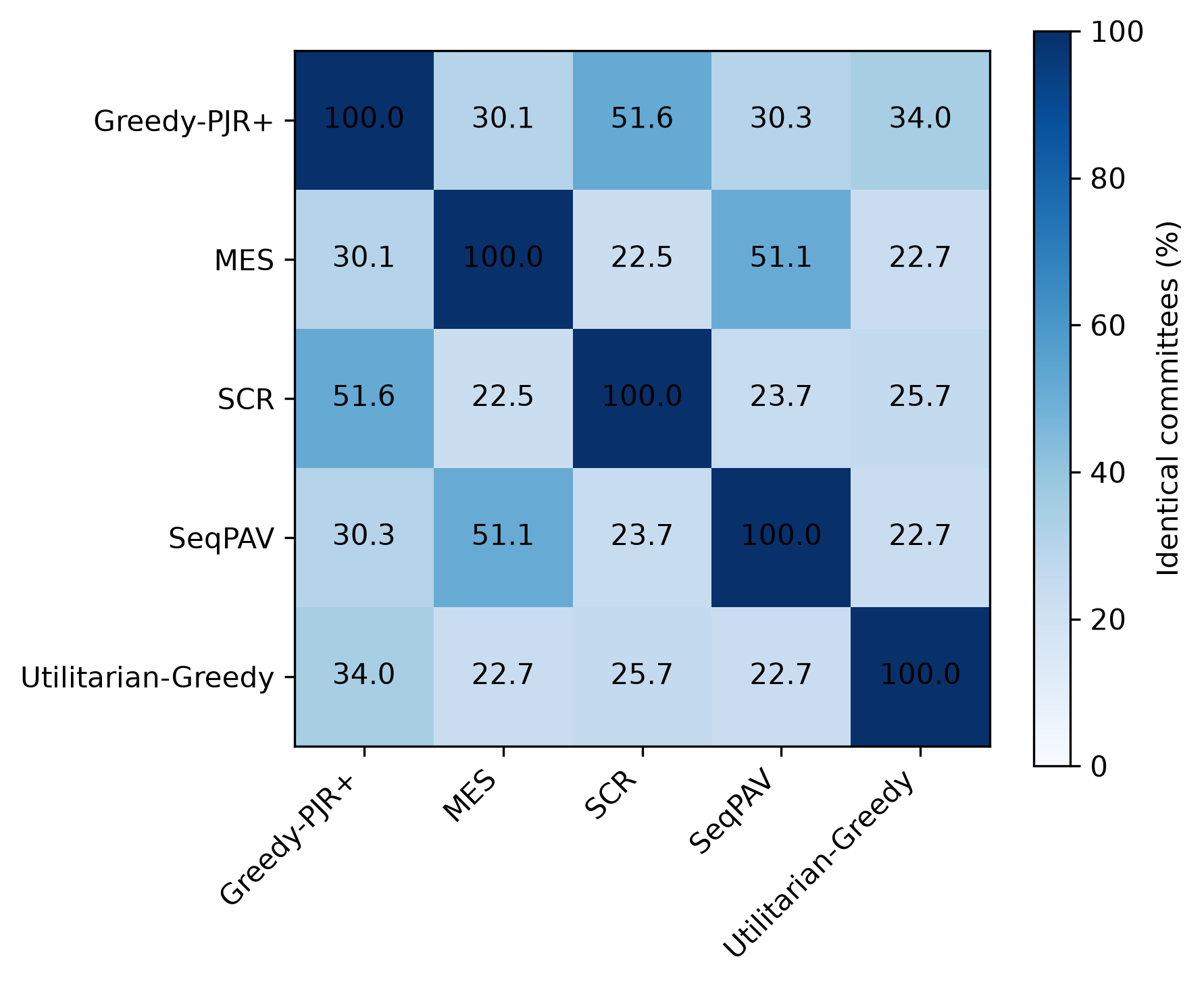}
				\end{minipage}
				\caption{Similarity of outcomes between the considered rules, in the ranked and approval setting, respectively.}
				\label{fig:similarity}
			\end{figure}

\begin{figure}[t]
				\centering
				\includegraphics[width=0.85\textwidth]{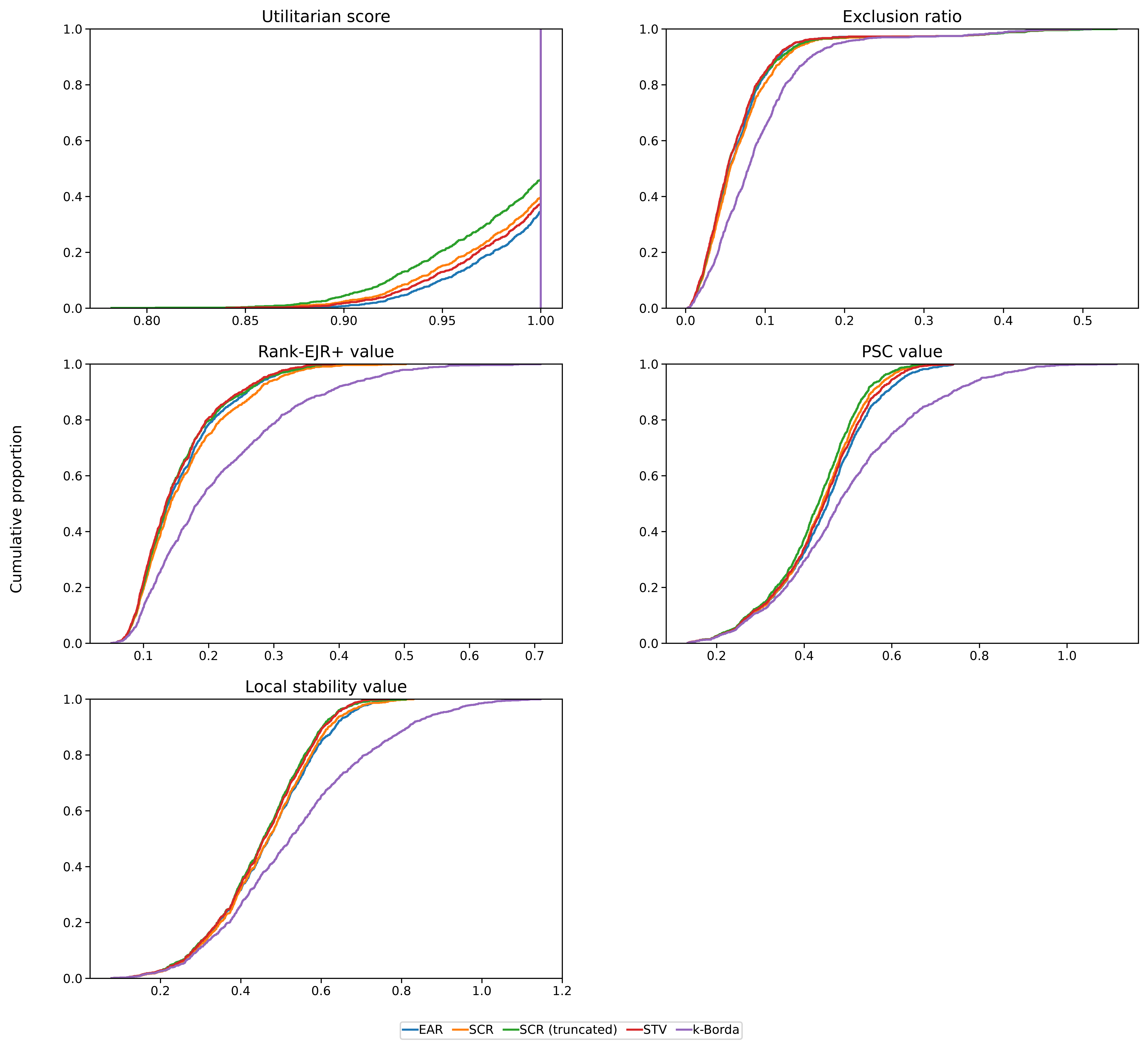}
				\caption{CDFs of the analyzed metrics for ranked ballots.}
				\label{fig:ranks_cdfs}
			\end{figure}

\begin{figure}[t]
				\centering
				\includegraphics[width=0.85\textwidth]{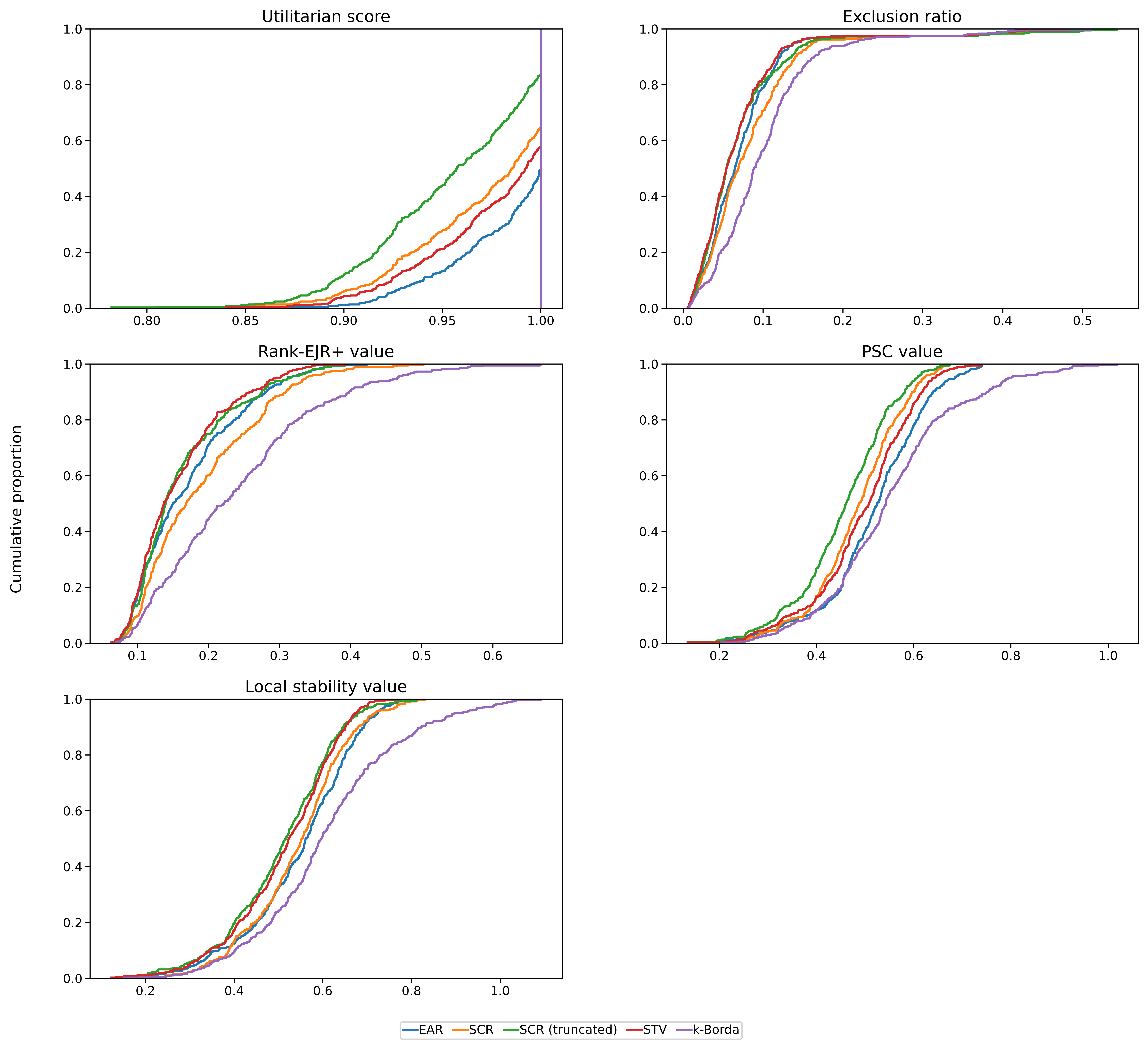}
				\caption{CDFs of the analyzed metrics for ranked ballots, restricted to instances where the considered rules return different committees.}
				\label{fig:ranks_cdfs_different}
			\end{figure}

\begin{figure}[t]
				\centering
				\includegraphics[width=0.85\textwidth]{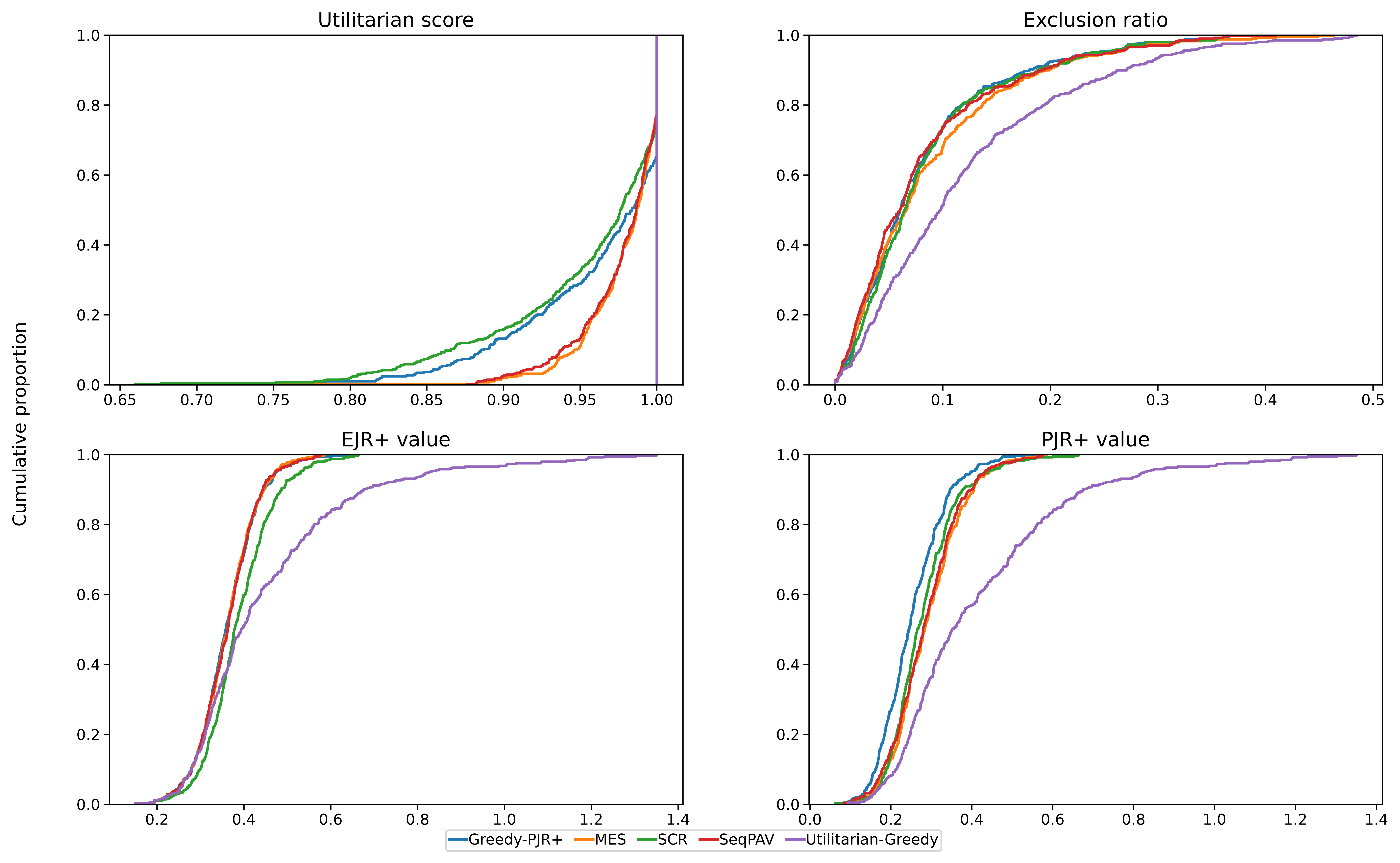}
				\caption{CDFs of the analyzed metrics for approval ballots.}
				\label{fig:approvals_cdfs}
			\end{figure}

			In what follows, we present the CDFs of the analyzed metrics across the two datasets used, which give a more nuanced and complete picture of the relationship among the considered rules; see \Cref{fig:ranks_cdfs,fig:approvals_cdfs}.
			We observed substantial similarities between the outcomes of the examined rules in the ranked-ballot setting (Scottish local elections dataset), likely due to the small size of the instances.
			In contrast, such similarities do not appear in the approval voting setting; see \Cref{fig:similarity}.
			For this reason, for the ranked-ballot setting, we additionally report CDFs restricted to instances where the considered rules produce different outcomes \Cref{fig:ranks_cdfs_different}, allowing for a more informative comparison.

			For the ranked-ballot setting, the Utilitarian score clearly differentiates the rules. EAR achieves the best performance, followed by STV, while both variants of SCR perform similarly and slightly worse.
			For Exclusion Ratio, STV performs best. EAR and SCR-Truncated achieve similar results, with slightly different characteristics: SCR-Truncated performs slightly better on average, but exhibits a heavier tail. SCR-Full performs worse than the other proportional rules, while $k$-Borda is, as expected, the worst-performing rule.
			For Rank-EJR+, STV, SCR-Truncated and EAR achieve the best results, while SCR-Full performs worse. Again, $k$-Borda is clearly the worst-performing rule.
			The PSC value is dominated by SCR-Truncated, followed by SCR. We note, however, that these rules are specifically designed to optimize this metric. STV performs next, followed by EAR, which in turn outperforms $k$-Borda.
			Finally, Local Stability is again best achieved by STV and SCR-Truncated, followed by SCR and EAR, although the differences between the latter two rules are very small. $k$-Borda is again significantly worse.

			For the approval voting setting, the Utilitarian score is maximized by MES and SeqPAV. After a noticeable gap, Greedy-PJR+ follows, with SCR performing very closely behind.
			For Exclusion Ratio, all proportional rules achieve comparable performance, with the main difference being the substantially worse performance of the Utilitarian rule.
			For EJR+ value, MES, SeqPAV, and Greedy-PJR+ achieve the best results, followed by SCR, while the utilitarian rule performs the worst.
			Finally, as expected, Greedy-PJR+ achieves the best PJR+ performance, with the Utilitarian achieving the worst. The remaining rules perform similarly.

		\end{document}